# Title: Carbon dioxide to carbon nanotube scale-up

**(Team C2CNT Round 2 Submission**
***extracts nrg cosia* CARBON XPRIZE)**

**Team Leader C2CNT: Prof. Stuart Licht**
Department of Chemistry, George Washington University
Washington, DC, 20052, USA

Abstract:
**Team C2CNT's (Carbon dioxide to carbon nanotubes) proprietary technology directly removes the widest range of carbon dioxide from the ecosystem. C2CNT technology simply transforms carbon dioxide into carbon and oxygen, and the carbon produced is permanently removed (stable on the order of geologic time frames). C2CNT technology directly removes, transforms and stores atmospheric (0.04%) $CO_2$, and directly removes 5% $CO_2$ is directly (without pre-concentration) removed from the air, up 5% CO2 (pertinent to the Track B removal of gas power plant $CO_2$ emissions of this submission). C2CNT technology directly removes, transforms and stores removes 5% CO2 is directly (without pre-concentration) removed from the air, 12.5% $CO_2$ (pertinent to the Track A removal of coal power plant $CO_2$ emissions of this submission), 33% $CO_2$ (pertinent to the complete removal of $CO_2$ from cement production plants), or 100%.**

Document:
The Carbon XPrize is an international competition to transform the carbon dioxide from fossil fuel plant emissions to the most valuable product (CarbonXprize.org). Track A is for the most valuable product from a coal plant emission (which includes not only the greenhouse gas CO2, but also ppm levels of $SO_2$, NOx and CO) and Track B is for the most valuable product from a natural gas power plant emission (which does not include the other pollutants). This document consists of Team C2NCT's Oct. 11, 2017 Round 2 Carbon XPrize Submission and Oct. 12, 2017 Addendum extended extracts).

C2CNT's mission:
Mining carbon dioxide to rehabilitate the planet.

The planet is becoming increasingly unsuitable to support human life due to greenhouse gas global warming. Climate change is also killing off entire species at an accelerating rate.

Which headline would you prefer?
Last year the average American generated 16 metric tons of $CO_2$.
or
Last year the average American generated 1 million dollars of carbon nanotubes **and no $CO_2$**.

**Team C2CNT Round 2 Submission Extracts**
***nrg cosia* CARBON XPRIZE**

**Submission Notes**

**Climate change and global warming are a growing threat to the continuity of life on the planet Earth. Climate change caused by increased greenhouse gas levels is killing, and is increasingly killing off (will be the direct cause of the extinction) of as much as 50% of the species on our planet by the turn of the century.**

**This is a report by the CARBON XPRIZE Team C2CNT as a Round 2 Submission of the development of the team's proprietary *technology to end global warming produced by the principal greenhouse gas carbon dioxide,* and remove anthropogenic carbon dioxide which has been released and growing at an alarming rate since the inception of the industrial revolution.**

**Team C2CNT's proprietary technology directly removes the widest range of carbon dioxide from the ecosystem. C2CNT technology simply transforms carbon dioxide into carbon and oxygen, and the carbon produced is permanently removed (stable on the order of geologic time frames). C2CNT technology directly removes, transforms and stores atmospheric (0.04%) $CO_2$, and directly removes 5% $CO_2$ is directly (without pre-concentration) removed from the air, up 5% CO2 (pertinent to the Track B removal of gas power plant $CO_2$ emissions of this submission). C2CNT technology directly removes, transforms and stores removes 5% CO2 is directly (without pre-concentration) removed from the air, 12.5% $CO_2$ (pertinent to the Track A removal of coal power plant $CO_2$ emissions of this submission), 33% $CO_2$ (pertinent to the complete removal of $CO_2$ from cement production plants), or 100%.**

**The C2CNT process is very low cost. Even more important, the C2CNT product is an extremely valuable form of carbon (carbon nanotubes). For example, carbon nanotubes are thousands of times more valuable than coal. The low production cost of carbon nanotubes by C2CNT technology is intended to substantially increase their market share to become the clear replacement product of choice for the billion dollar steel and aluminum (due to CNT's much high strength to weight ratio) for products as varied as car bodies to lighter-weight laptop computers. Carbon nanotube's outstanding electrical and thermal conductivity, torsional and blast resistance make them preferred additives to improved, ceramic, cement and opens a wide range of new textiles, such as veneer thin, bullet and taser-proof suits.**

**C2CNT provides a major economic incentive to "mine" this greenhouse gas and change carbon dioxide from pollutant to desired resource**

**C2CNT TEAM DETAILS**

**Address: George Washington University, Science & Technology campus, Ashburn, VA , USA**

**C2CNT TEAM LEADER**

**Prof. Stuart Licht, Dept. of Chemistry George Washington University**

**ADDITIONAL C2CNT TEAM MEMBERS SINCE THE JULY 27, 2016 INCEPTION OF THE CARBON XPRIZE**

**Mathew Lefler, Dept. of Chemistry George Washington University**
**Juan Vicini, Dept. of Chemistry George Washington University**

**OTHER CURRENT, LONG TERM C2CNT TEAM MEMBERS**

**Dr. Marcus Johnson, Dept. of Chemistry George Washington University**
**Brian Wright, Dept. of Chemistry George Washington University**

**PAST AND PRESENT MEMBERS WHO HAVE PARTICIPATED FOR SHORTER PERIODS IN THE C2CNT TEAM**

**Dr. Jason Lau, Dept. of Chemistry George Washington University**
**Dr. Richa Singhal, Dept. of Chemistry George Washington University**
**Dr. Jiawen Ren, Richa Singhal, Dept. of Chemistry George Washington University**
**Xinye Liu, Dept. of Chemistry George Washington University**
**Gad Licht, Dept. of Chemistry George Washington University**
**Parth Contractor, Dept. of Chemistry George Washington University**
**Zhikun Wang, Dept. of Chemistry George Washington University**
**Mohamed Eltahir, Dept. of Applied Economics and Sustainability**
**Azhar Elahir, Dept. of Chemistry George Washington University**
**Zijun Shao, Dept. of Chemistry George Washington University**

ABOUT US: The C2CNT team focus is a comprehensive solution to climate change, a principal challenge facing the planet today. In this industrial age, massive quantities of greenhouse gases, principally carbon dioxide, have been released into our atmosphere through the combustion of fossil fuels to meet humankind's industrial, transportation and energy needs. These greenhouse gases trap infrared heat increasing the temperature of the planet (global warming) at an accelerating rate. Climate change causes species extinction. As many as half the species on the planet face extinction this century if climate change is not abated. Climate change is causing sea-level rise, which is already threatening disappearance of low-lying countries, glaciers and icebergs, along with causing interior flooding, drought, famine, disease, hurricanes, increased incidence and strength of tornados, communication, power and transportation disruptions, and causing wide-spread increasing economic cost to both developed and under developed countries throughout the world.

At George Washington University we've discovered the inexpensive transformation of the greenhouse carbon dioxide into a widely useful and highly valued product. In our C2CNT process, $CO_2$ is directly transformed to hollow nanofibers, "carbon nanotubes," products with remarkable properties of conductivity, nanoelectronics, higher capacity batteries, flexibility, with greater strength than steel and widespread use as carbon composites. The conversion of $CO_2$ to pure carbon nanotubes provides the most compact form to capture carbon dioxide and mitigate climate change. The market for carbon composites provides lighter weight alternatives to metals, and is used today in the Boeing Dreamliner, high end sport cars, and athletic equipment. The market is experiencing an explosive growth comparable to the historical start of the plastic industry. Previously carbon nanofibers were made by expensive processes such as chemical vapor deposition or polymer pulling and could not be made from $CO_2$. Our C2CNT team is rapidly scaling-up our new chemistry which directly converts $CO_2$ at high rate to carbon nanotubes using low cost materials. The C2CNT team is committed to reversing the rapid anthropogenic buildup of greenhouse gases, and perceive $CO_2$ not as a pollutant, but rather $CO_2$ as a useful resource. We are working to incentivize $CO_2$ removal from the power, industry and transportation sectors by rapidly and efficiently transforming $CO_2$ into a valuable product.

**ABOUT C2NCT TEAM LEADER: Prof. Stuart Licht**

Team Leader Bio: The C2CNT path started in 1987 with Stuart Licht's theory and experiment for the efficient use of solar energy to drive simultaneous electrical and chemical production [1] and new room and high temperature electrochemistries to store energy [2]. The C2CNT process is the culmination of 30 years ongoing development of his technical solution to a sustainable future. The process matured to a new theory [3a] & method [3b] of solar energy conversion to enhance of splitting water to fuel by using the full spectrum of sunlight (visible & thermal) and electrochemistry to greatly increase the solar energy conversion to hydrogen fuel efficiency. This led to a variety of new molten salt electrolysis chemistries to produce societal staples such as fertilizer, cement and metals including iron, but without the $CO_2$ emissions normally associated with those processes [4]. In 2009, the theory of solar thermal electrochemical process (STEP) water splitting was extended to carbon dioxide splitting [5a], leading to the experimental demonstration in 2010 of $CO_2$ splitting at > 50% solar energy efficiency [5b], and the subsequent demonstration in molten salts of the direct transformation of $CO_2$ to variety of carbon, plastics precursors, and synthetic methane fuels [5c. Such products made from $CO_2$ are valued at only ~$40 to $400 per ton and therefore give little financial impetus to the removal of this greenhouse gas. Starting In 2012 – 2013 Licht realized that an orders of magnitude more valuable product, carbon nanotubes, could be produced by direct molten carbonate electrolysis of $CO_2$ leading to the new C2CNTprocess, and providing a significant economic imperative to the removal of this anthropogenic greenhouse gas. Today, the team leader is a George Washington University Chemistry Prof. He previously served as a Program Director at the National Science Foundation, and received awards including the Hildebrand Chemical Prize (2016), The BASF Energy Storage Prize in honor of BASF's 150th anniversary (2015), the Electrochemical Society Energy Award (2006) and the Alcoa Science Prize (1994).

1. Licht, *Nature*, cover article, *330*, 148, 1987; Licht et al, *Nature*, *326*, 863, 1987; *Nature*, *345*, 330, 1990; *Nature*, *354*, 440, 1991.
2. Licht et al, *Science, 261*, 440, 1991; *Science, 285*, 1039, 1999; *Energy & Environ Chem*, *6*, 3646*, 2013*; *ACS Central Sci*, *2*. 162, 2016.
3. a. Licht, *EChem Com*. 4623, 2002; Licht, *J Phys Chem*, 4243, 2003. b. Licht et al, *Chem Comm*, 3006, 2003.
4. Licht et al. *Science*, , *345*, 639, 2014, Chem Comm, 7004; 2010; 2081, 2011; 6019, 2012,
5. a. Licht, *J Phys Chem C*, 2009. b. Licht et al, *J Phys Chem Lett*, 2010. c. Licht et al., *Adv. Mat*., 2011; *Adv. Energy Mat*., 2015; *Adv. Science*, 2015; *Adv. Mat. Tech.*, 2016.

**Team Video (YouTube URL):** https://www.youtube.com/watch?v=gVuyJqsEQxw&feature=youtu.be

**C2CNT IN THE NEWS**

C2CNT is the Licht group's low energy, low cost electrochemical pathway to transform $CO_2$ to carbon nanotubes. For the first time a very high value product is made at low cost to incentivize the greenhouse gas removal. This chemical process evolved from prior Licht group chemical development of the solar driven molten salt electrolysis of carbon dioxide. Presentation of a comprehensive solution to global warming by the electrolysis of $CO_2$ to a valuable product is delineated in the Tech Information section, as well as in C2CNT Popular & Technical Media C2CNTdescriptions such as:

•Licht interview on BBC World Newshour; click on the audio button at "The Bottom of the Barrel?" This will start the Newshour, and slide ahead to the interview between time 18:52 and 23:00.14. http://www.bbc.co.uk/programmes/p02yxbv0

•description in *Science* http://science.sciencemag.org/content/349/6253/1158.full

•description in *Forbes* http://www.forbes.com/sites/ericmack/2015/08/19/science-turns-climate-change-gases-into-planes-better-batteries-much-more/#7b732de4351e

•description in *MIT Technology Review* https://www.technologyreview.com/s/600939/how-carbon-dioxide-from-the-air-can-boost-batteries/

•Press conference at American Chemical Society National Meeting https://www.youtube.com/watch?v=kQBlpENiFzo

•description of C2CNT at phys.org http://phys.org/news/2016-06-power-co2-emissions-carbon-nanotubes.html

•Licht C2CNT interview on public television (scitech)now episode 218 at:
The introduction, interview & Licht's $CO_2$ capture animation runs from 00:00-00:34 and 13:44-19:22
http://www.scitechnow.org/videos/episode-218/Year three milestone: Global deployment, starting a C2CNT 10 yr plan to lower atmospheric $CO_2$ to < 400 ppm.

ROUND 2 TECHNICAL DETAILS

SCIENCE, ENGINEERING AND PROCESS OVERVIEW
The C2CNT process directly transforms the carbon dioxide greenhouse gas in an electrolyzer chamber into a particularly valuable hollow core form of carbon nanofibers (CNFs) called carbon nanotubes (CNTs) plus oxygen gas. As used in this CCEMC project, C2CNTis utilized to eliminate all emissions from a natural gas (NG) power plant. The electrolyzer's oxygen gas is looped back into the power plant to increase combustion efficiency. The carbon nanotubes can provide an excellent return on investment as they are inexpensively formed from the hot flue gas, and are valued at up to 1000 fold higher than fuels or other carbon products made from the carbon dioxide. The high value product is useful for a range of applications including as lightweight replacement to the trillion dollar metals market (as well as non-metal construction and packaging materials, batteries and nanoelectronics), provides economic incentive to eliminate the greenhouse gas, and compactly store the $CO_2$ as useful, pure carbon. In particular, the CNTs can address the growing scarcity of the world's metals offering the world a nearly endless supply of this stronger, lighter, electrically and thermally conductive material.

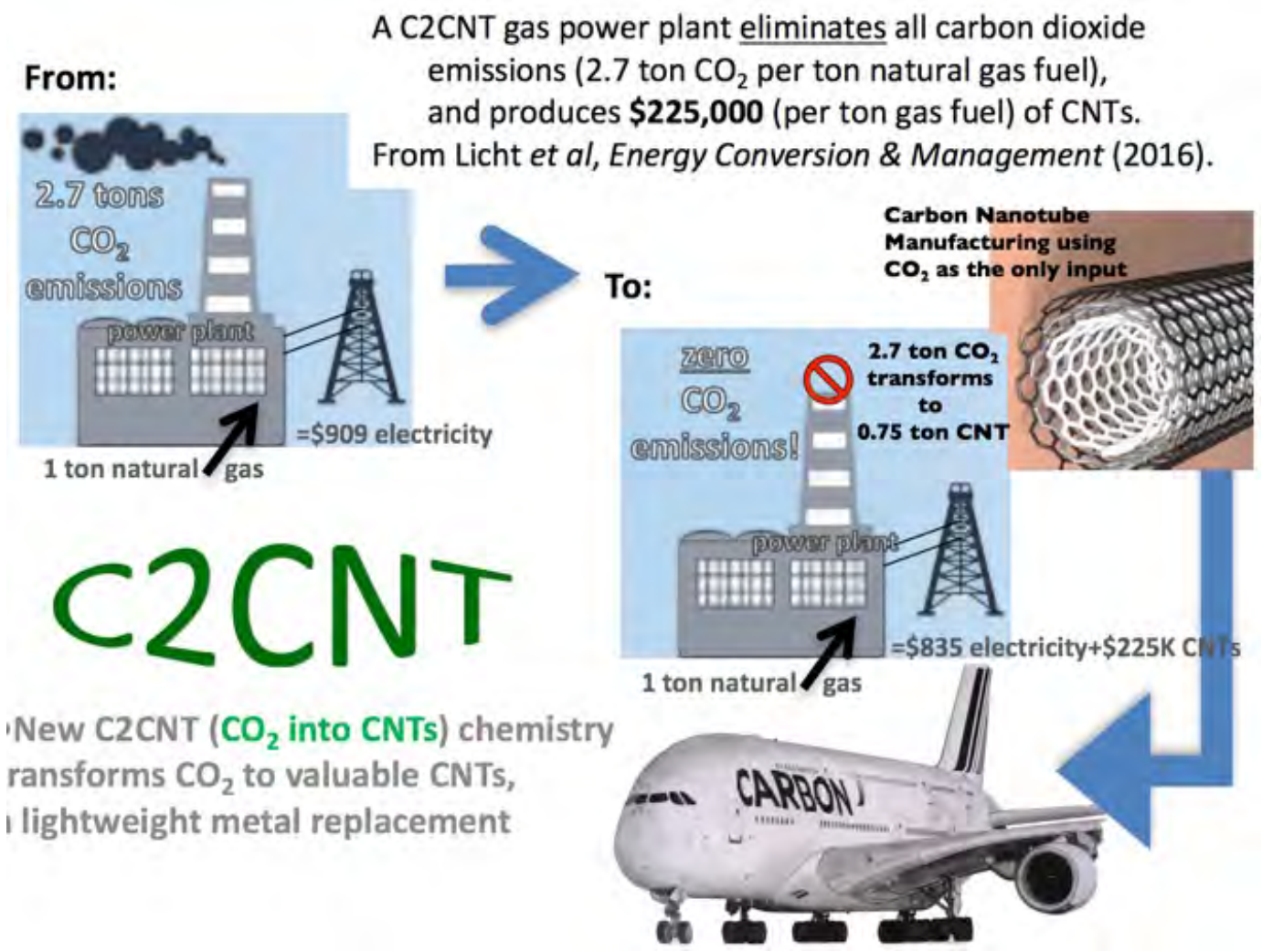

The ongoing consequences of increased levels of the greenhouse gas carbon dioxide include species extinction, climate change and famine. Carbon dioxide elimination is incentivized by its transformation into a valuable product. The team is rapidly scaling up the $CO_2$ to carbon nanotube process, and the Round 2 CCEMC project is planned to convert 5 tons (metric)/day of $CO_2$ emissions at an NG gas combined cycle (CC) power plant in Alberta into valuable carbon nanotubes to decrease this greenhouse gas and to ameliorate climate change. CNTs have superior strength, conductivity, flexibility and durability, but were limited by their high reactant & production costs. The team discovered the inexpensive electrolytic transformation of $CO_2$ from flue gas dissolved in molten carbonates into CNTs & $O_2$. $O_2$ improves NG CC power plant efficiency. As an example per ton NG fuel:

without C2CNT, plant output: 2.7 ton $CO_2$, $909 electricity & $0 of CNT;
with C2CNT, plant output: 0.0 ton $CO_2$, $835 electricity & $225,000 CNT.

$CO_2$ elimination is incentivized by its transformation into a valuable product. In the C2CNT process, the carbonate electrolyte is not consumed and the net reaction is $CO_2$ splitting into carbon and $O_2$, as presented here using pure $Li_2CO_3$ as the carbonate electrolyte:

Dissolution: $CO_2(gas) + Li_2O(soluble) \rightarrow Li_2CO_3(molten)$
Electrolysis: $Li_2CO_3(molten) \rightarrow C(CNT) + Li_2O(soluble) + O_2(gas)$
Net: $CO_2(gas) \rightarrow C(CNT) + O_2(gas)$ (1)

Note from the net eq 1, that all $CO_2$ gas is transformed carbon nanotubes (CNTs), and that from their respective molecular weights, 1 ton of CNT is formed per 3.7 ton of $CO_2$ consumed. CNTs are stronger than steel, stable and compact providing an ideal means to remove, transform and store $CO_2$ from flue gas.

The C2CNT electrolyzer addition to the existing NG CC power plant in Alberta will be built on site, using materials (nickel, steel, lithium carbonate); the CNTs produced from the transformed 5 ton per day of $CO_2$., instead of the $CO_2$ being released as a greenhouse pollutant._The C2CNT electrolyzer addition to the existing coal power plant in Wyoming will be built on site, also using materials (nickel, steel, lithium carbonate); the CNTs produced from the transformed 5 ton per day of $CO_2$.

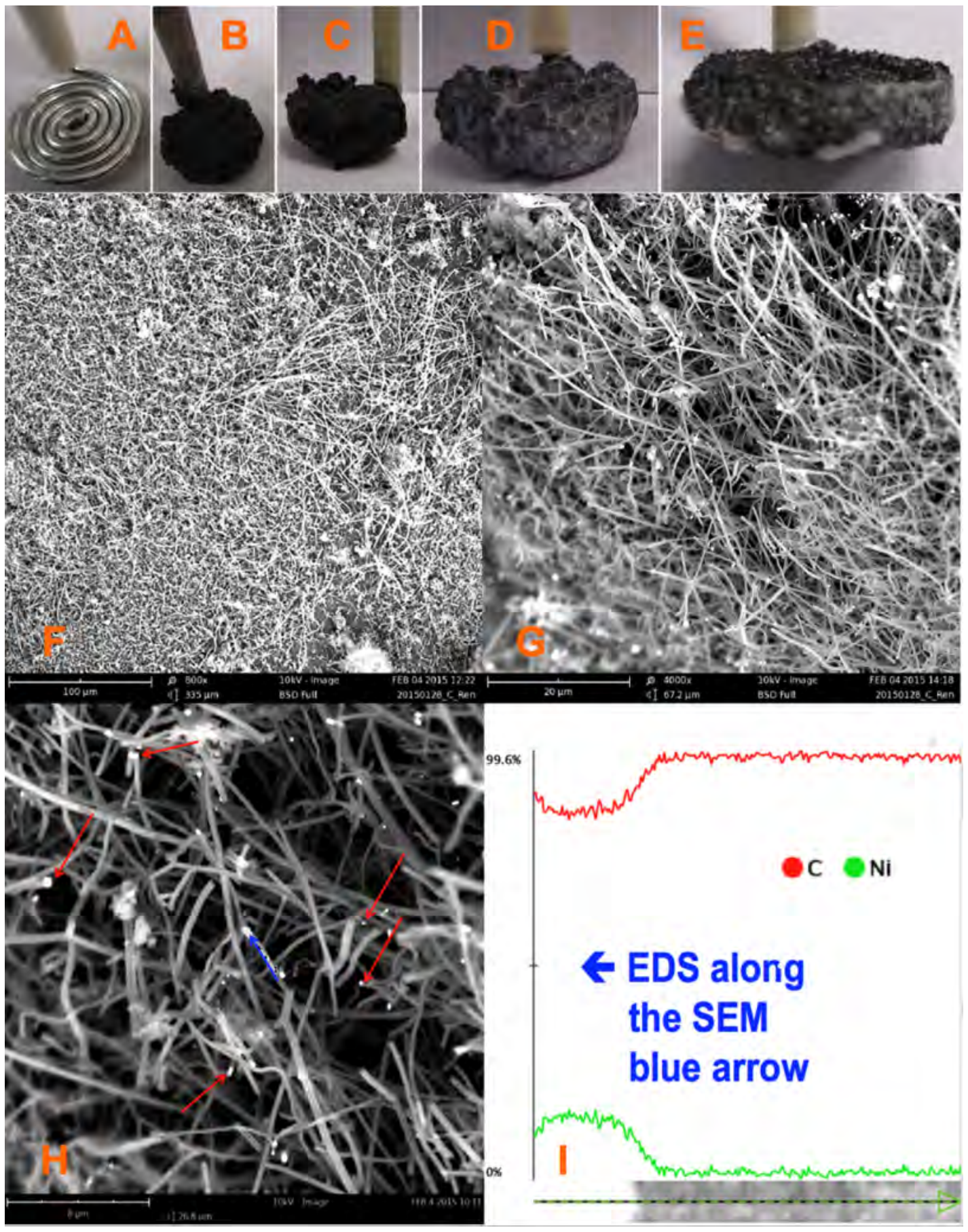

Carbonate's higher concentration of active, reducible tetravalent carbon sites logarithmically decreases the electrolysis potential and can facilitate charge transfer at low electrolysis potentials. We observe that carbonate is readily split to carbon approaching 100% coulombic efficiency (coulombic efficiency is determined by comparing the moles of applied charge to the moles of product formed, where each mole of solid carbon product formed depends on four moles of electrons). C2CNT's carbon nanotubes made from carbon dioxide have a strength greater than steel, and can have a wide variety of morphologies, activities and battery storage capacities. C2CNT has discovered how to make uniform syntheses of each of these CNT types as delineated in a wide range of our recent, attached scientific articles. Examples include very long length CNTs useful for weaving into textiles (google Licht Carbon Nanotute Wool), highly conductive boron doped carbon nanotubes and twisted CNTs with a particularly high battery energy storage capacity.

$CO_2$ to carbon nanotubes are formed at a simple metal cathode and a simple metal cathode The blue arrow originates at one Ni site and moves along the CNF path. "I": EDS composition mapping along the 6µ blue arrow path shown in SEM "H".

C2CNT Round 2 occurs at only one site

TWO TRACK RATIONALE AND DETAILS
C2CNT Technology, which directly transforms the greenhouse gas $CO_2$ to carbon nanotubes, (I) functions well at the 0.04% (ii) concentration from the atmosphere up through pure (100%) $CO_2$, and hence is equally well situated to transform and remove intermediate concentrations of $CO_2$ such as either Track A's 12.5% or Track B's 5% CO2. (ii) As we reported and published (and as described in the attachments) C2CNT is highly impervious to the ppm level concentration SO2, NOx or CO levels found in Track A's emission. (ii) The C2CNT transformation of carbon dioxide to CNTs is impervious to severaly percent of SO2, CO or NOx (and can lead to useful CNTs with useful, specialized properties at even higher concentrations of these pollutants as delineated in our recent, attached scientific articles. Hence, the Track A synthesis required with these syntheses will not be affected by a Track B synthesis which does not require, but with C2CNT can include, these pollutants.

Inputs and Outputs
Inputs: $CO_2$ and electricity; outputs: carbon nanotubes and oxygen

PRODUCT DEFINITION
Multiwalled Carbon Nanootubes are rolled, concentric cylinders of graphene (single layers of $sp^2$ bonded carbon), which are unusually strong, but had previously been expensive to proeuce.

PROCESS STOICHIOMETRY:
$CO_2 \rightarrow C_{CNT} + O_2$

TECHNICAL DOCUMENTATION

EXPLANATORY NOTES

The C2CNT electrolyzer feature in either Track A coal flue gas emission and Track B gas flue gas emission, the C2CNT electrolyzer feature remains the same and functions in a similar manner. In both cases, gases other than carbon dioxide, such as a nitrogen, oxygen and water vapor or insoluble in the molten carbonate electrolyte which is used to both dissolve carbon dioxide and split, by electrolysis, the dissolved carbon dioxide into valuable carbon at one electrode and pure oxygen at the other electrode. In each case the pure oxygen can be looped back into the inlet of the combusting coal or gas process (the oxy-fuel advantage) which provides the capability to improve the efficiency of each of the power plants. As illustrated below, we have designed and analyzed fully functional C2CNT gas and coal plants. The gas plant simply emits flue gas without carbon dioxide, while the coal plant utilizes a renewable energy boost to drive the C2CNT process without impacting the electrical output of the coal plant. In each case the value of the carbon nanotubes far exceeds the value of the electricity produced to provide considerable impetus to “mine/remove/recover” and prevent all carbon dioxide from entering the environment.

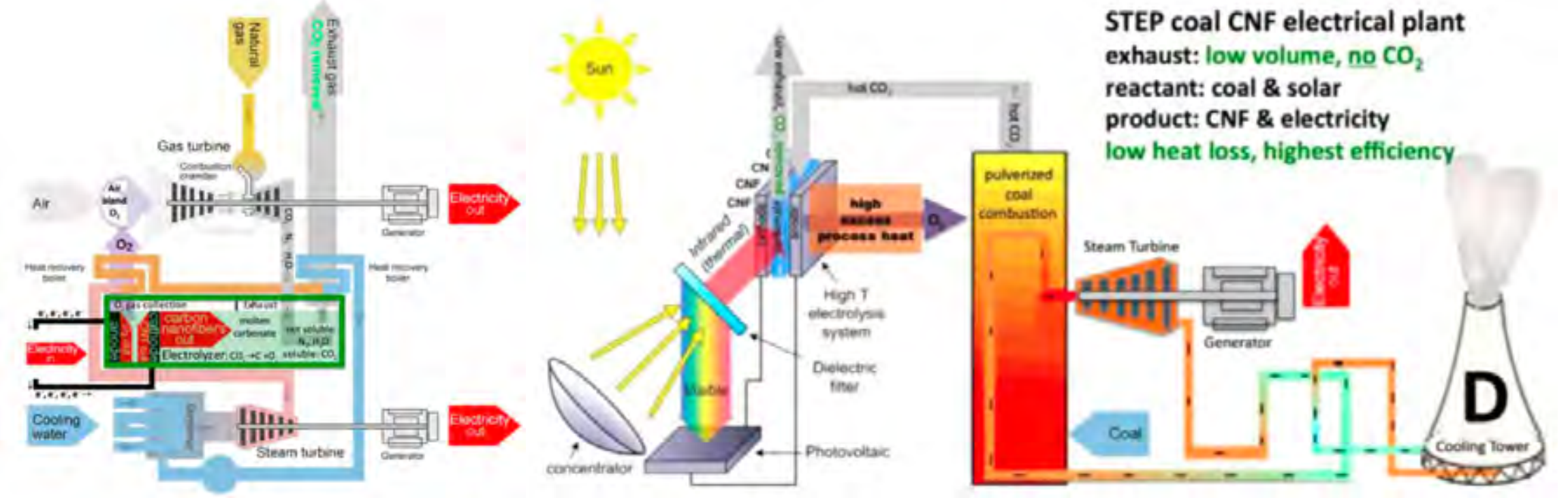

Top: A C2CNT combined cycle natural gas power plant. Bottom: A C2CNT coal power plant with renewable energy “STEP (Solar Thermal Electrochemical Process) assist.

Bottom Left: C2CNT PROCESS FLOW DIAGRAM; the single system combines both capture and conversion systems. Bottom Right: PIPING & INSTRUMENTATION DIAGRAM; the single system combines both capture and conversion systems. Note that low levels of additional pollutants, such as NOx, $SO_2$ and CO in conventional coal power plants are notshown and do not affect the carbon dioxide to carbon nanotube transformation.

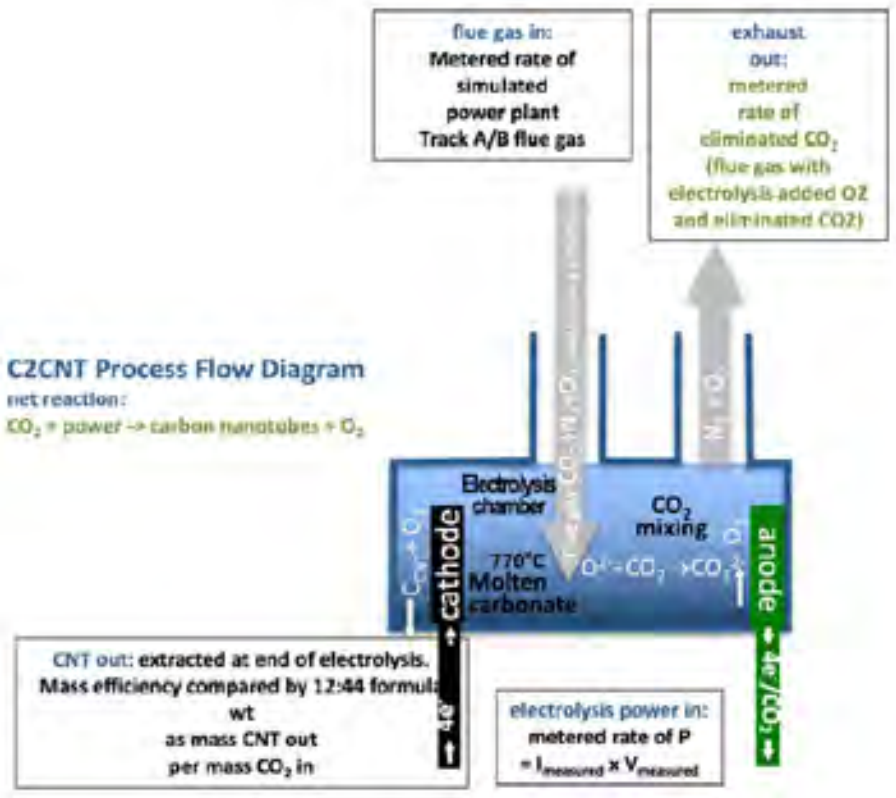

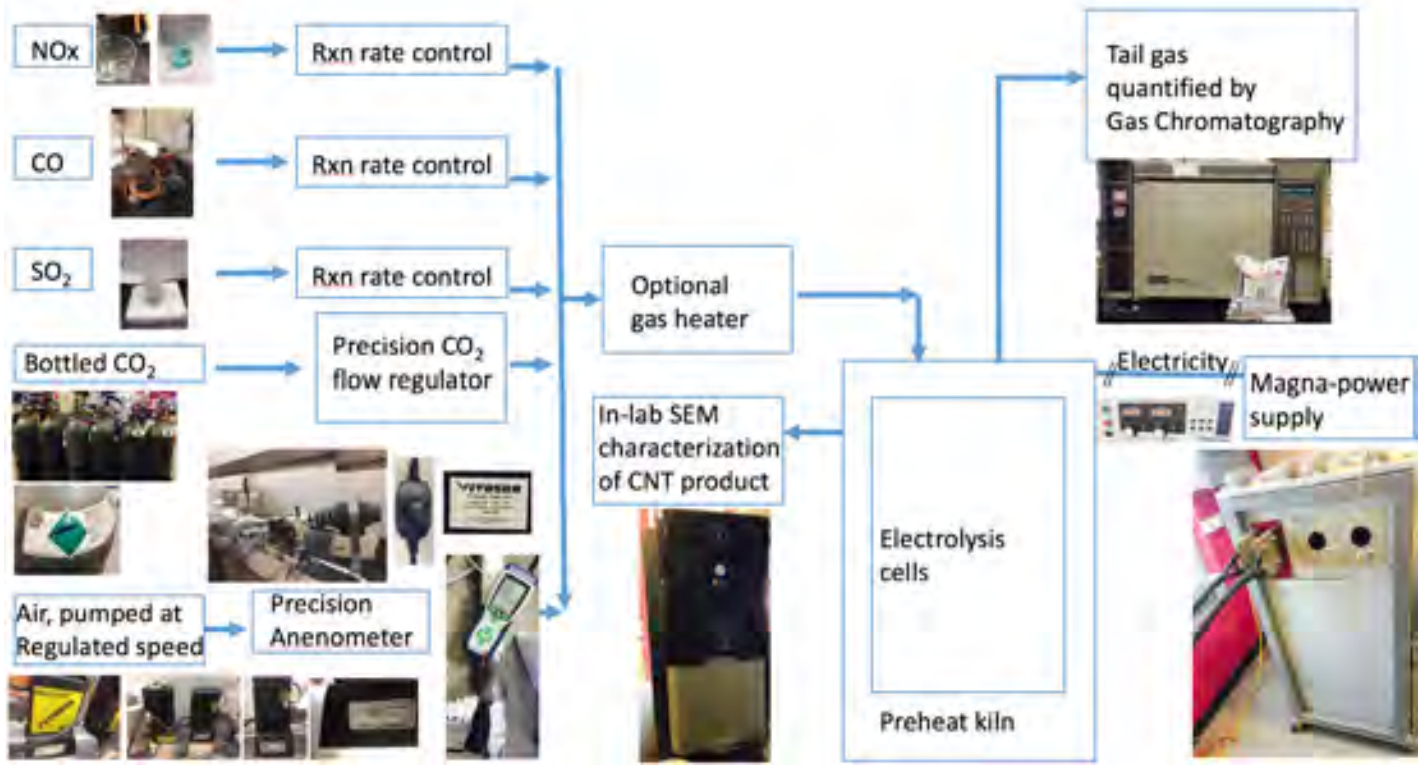

## MASS AND ENERGY BALANCE

A full description of the C2CNT Energy and Mass Balance is given in sections 3-7 of the C2CNT 2016 paper

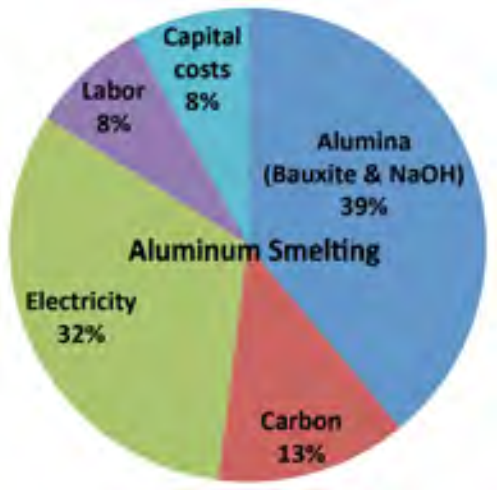

"Thermodynamic assessment of $CO_2$ to carbon nanofiber transformation for carbon sequestration in a combined cycle gas or a coal power plant" (attached) with the pertinent features that all carbon dioxide is consumed and yields a net of only two products (carbon nanotubes an pure oxygen), that only a small fraction of the electricity is consumed in the electrolysis and that useful heat is liberated in the dissolution of carbon dioxide in carbonate. In addition, an abbreviated mass and energy balance here is derived from a comparison to that known of a comparable, mature industry: aluminum production. The C2CNT process bears many similarities to aluminum smelting. Both processes consist of molten electrolysis, and do not use noble or exotic materials. Aluminum smelting produces aluminum metal from alumina (using bauxite, sodium hydroxide and electricity), while C2CNT produces carbon nanotubes from carbon dioxide (using carbon dioxide and electricity). Aluminum smelting operates at 960°C in a molten cryolite electrolyte. The C2CNT process operates under somewhat milder conditions at 770°C in a less exotic, molten carbonate electrolyte. Both processes operate at high rate (hundreds of mA per $cm^2$) and low polarization. In both cases the electrolysis chamber consists of common metals, common insulators (such as kiln or "firebricks"), and control equipment. Electrolysis in the Aluminum smelting process is driven at approximately 4 volts using 3 electrons per aluminum. The C2CNT electrolysis is driven at approximately 1 volt using 4 electrons per carbon dioxide. C2CNT and aluminum smelters have approximately equivalent output (tonnage) rates. Aluminum costs ~$1,880 per metric ton, of which ~32% of the cost is electricity (Djukanovic, 2012). Today's newer, more efficient Al plants require 12 MWh per ton, whereas older Al plant require 15 MWh per ton. 13 MWh is measured and calculated from a 94% efficient 3 electron per Al electrolysis at 4.1V (Naixiang, 2014). $600 for 12 MWh = $0.05 per kWh. This Al electricity cost varies from lowest (in the Middle East) at $300, to mid range $650 (US), to highest (China with high energy tariffs) of $1,020/ton. Al smelter cost structure (modified from Djukanovic, 2012). Production costs per metric ton (tonne) of Al, based on market costs consist of: Consumable Expenses (32% electricity & 52%; reactants = 84%), Electricity: 32%, Labor: 8%, and Capital Expenses (amortized cost of electrolyzers, processing equipment, and miscellaneous overhead). Note that the energy to drive the aluminum production originates from two sources (electricity and energy released from the consumed carbon anode).

Table 1. Comparison of aluminum and C2CNT production costs.

| Process | $US Cost (% of total) | | | | | |
|---|---|---|---|---|---|---|
| | alumina | carbon | electricity | labor | capital | total |
| Aluminum | 733(39%) | $244(13%) | $602(32%) | $150 (8%) | $150 (8%) | $1880(100%) |
| | $CO_2$ | - | electricity | labor | capital | total |
| C2CNT | $0 | $0 | $360 | $150 | $150 | $660 |

As compared in Table 1, and unlike Al smelting, the C2CNT process uses a no-cost oxide as the reactant (carbon dioxide, rather than bauxite). Both are straightforward, high current density electrochemical (molten electrolytic reduction of oxide) processes. The C2CNT process operates under somewhat milder conditions at 770°C in a less exotic, molten carbonate electrolyte at similar rates of output, and to a first order of approximation, both processes will be assumed to have the same labor and capital costs. Whereas, Al production requires ~13 MWh per ton of aluminum, C2CNT production requires less energy (7 MWh) per ton of carbon nanotubes. This energy is calculated here from the C2CNT 1 volt electrolysis consuming 4 electron per carbon dioxide splitting efficiency. The observed electrolysis voltage varies from 0.8V to up to 2V, decreasing with higher concentrations of added lithium oxide, and increasing with current density and with mixed molten carbonate electrolytes (Ren, 2015; Ren, 2017). Using the formula weight to convert mass carbon dioxide to moles, and Faradays constant at 1V yields 2.4 MWh per ton $CO_2$, which decreases to 2.0 MWh per ton CO2 (which is 7.2 MWh per CNT) by the 20% energy recovered through driving the turbine more efficiently with pure oxygen (looped in from the C2CNT electrolysis), rather than regular air, combustion (Lau, 2016). This yields an electrical cost of $360 per ton CNT, and as summarized in Table 1, a total cost of $660 per ton of CNT. The electrical cost falls per ton CNT based on less expensive wind electric, equivalent to (x12.01/44.01) $50 per ton of $CO_2$ (Licht, 2017). Higher production rates will increase this cost, while the imposition of a carbon tax or carbon credits will lower this cost.

# Team C2CNT Round 2 Submission - Oct. 12, 2017 Addendum
# *nrg cosia* CARBON XPRIZE

### Addendum to Performance Data (Sect. 4 of the 25 page C2CNT written Submission)

The C2CNT team Carbon XPrize submission is unusual in the large amount of available, published, rigorously reviewed scientific articles and patents available documenting and supporting the Sect. 4 Performance Data of the 25 page C2CNT written Submission.

Here, in addition to the previously submitted attachments consisting of (1) the scientific articles, and (2) a list of the patents and publications distributed among the Team Leader's CV, the Team C2CNT includes the following information:

### Team C2CNT Round 2 Oct. 12, 2017 Addendum - *nrg cosia* CARBON XPRIZE
### Table of Contents

## 2015-2017 Bibliographic details relevant to Team C2CNT published articles

1. Marcus Johnson, Jiawen Ren, Matthew Lefler, Gadi Licht, Juan Vicini, Xinye Liu and Stuart Licht,
**"Carbon Nanotube Wools Made Directly from $CO_2$ By Molten Electrolysis: Value Driven Pathways to Carbon Dioxide Greenhouse Gas Mitigation**,"
***Materials Today Energy,* 5,** 230-236 (2017).

2. Marcus Johnson, Jiawen Ren, Matthew Lefler, Gadi Licht, Juan Vicini, Yinwe Liu and Stuart Licht,
**"Data on SEM, TEM and Raman Spectra of Doped, and Wool Carbon Nanotubes Made Directly from $CO_2$ by Molten Electrolysis,"** ***Data in Brief,*** published Aug. 17, (2017)**.**
**Available at:** http://www.sciencedirect.com/science/article/pii/S2352340917303906

3. Stuart Licht,
**"Co-Production of Cement and Carbon Nanotubes with a Carbon Negative Footprint,"**
***J. CO2 Utilization***, **18**, 378-389 (2017). March, 19. 2017. doi.org/10.1016/j.jcou.2017.02.011
http://www.sciencedirect.com/science/article/pii/S2212982016302852

4. Jiawen Ren, Marcus Johnson, Richa Singhl, Stuart Licht,
**"Transformation of the greenhouse gas $CO_2$ by molten electrolysis into a wide controlled** selection of carbon nanotubes," ***J. CO2 Utilization***, **18**, 335-344 (2017). March, 19. 2017.
http://www.sciencedirect.com/science/article/pii/S2212982016302864

5. Jason Lau, Gangotri Dey, Stuart Licht
**"Thermodynamic assessment of $CO_2$ to carbon nanofiber transformation for carbon sequestration in a combined cycle gas or a coal power plant,"**
***Energy Conservation and Manangement,* 122*,*** 400-410 (2016). available 6/7/16
open access at: http://authors.elsevier.com/sd/article/S0196890416304861

6. Stuart Licht, Anna Douglas, Jiawen Ren, Rachel Carter, Matthew Lefler, and Cary L. Pint
**"Carbon Nanotubes Produced from Ambient Carbon Dioxide for Environmentally Sustainable Lithium-Ion and Sodium-Ion Battery Anodes,"**
***ACS Central Science,* 2,** 162-168 DOI: 10.1021/acscentsci.5b00400 (2016).
available open access at:http://pubs.acs.org/doi/pdf/10.1021/acscentsci.5b00400

7. Hongjun Wu, Zhida Li, Deqiang Ji, Yue Liu, Lili Li, Dandan Yuan, Zhonghai Zhang, Jiawen Ren, Matthew Lefler, Baohui Wang, and Stuart Licht,
**"One-Pot Synthesis of Nanostructured Carbon Material from Carbon Dioxide via Electrolysis in Molten Carbonate Salts,"** ***Carbon,* 106,** 208-217 (2016).
available open access 5/14/16 at: http://www.sciencedirect.com/science/article/pii/S0008622316303852

8. Gangotri Dey, Jiawen Ren, Tarek El-Ghazawi, Stuart Licht
**"How does amalgamated Ni cathode affect Carbon Nanotube growth? A density functional theory study,"** ***RSC Advances*, 6,** 27191-27196 (2016).

9. Jiawen Ren, and Stuart Licht,
**"Tracking airborne $CO_2$ mitigation and low cost transformation into valuable carbon nanotubes,"**
***Scientific Reports,*** 6, 27760-1-11 (2016). available open access at:
http://www.nature.com/srep/2016/160609/srep27760/full/srep27760.html

10. Jiawen Ren, Fang-Fang Li, Jason Lau, Luis Gonzalez-Urbina, Stuart Licht,
**"One-pot synthesis of carbon nanofibers from $CO_2$,"**
***Nano Letters*, 15,** 6142-6148 plus 13 page published SI, (2015).

11. Jiawen Ren, Jason Lau, Matthew Lefler, Stuart Licht,
**"The minimum electrolytic energy needed to convert carbon dioxide by electrolysis in carbonate melts,"**
***Journal of Physical Chemistry, C*, 119,** 23342-23349 (2015).

**2015 to 2017 additional publications demonstrating lower value (than carbon nanotube) products**

12. Hongjun Wu, Deqiang Ji, Lili Li, Dandan Yuan, Yanji Zhu, Baohui Wang*,
Zhonghai Zhang* and Stuart Licht,
**"Efficient, high yield carbon dioxide and water transformation to methane by electrolysis in molten salts," *Advanced Materials Technology,* 1,** 60092 (2016)**.** ***online July 14, 2016***
DOI: 10.1002/admt.201600092 **(2016); July 14, 2016 at:**
http://onlinelibrary.wiley.com/doi/10.1002/admt.201600092/abstract

13. Stuart Licht, Shuzhi Liu, Baochen Cui, Jason Lau, Liwen Hu, Jessica Stuart, Baohui Wang,
Omar El-Gazawi, and Fang-Fang Li
**"Comparison of alternative molten electrolytes for water splitting to generate hydrogen fuel,"**
***J. Electrochem. Soc,* 163,** F1163-F1168-23349 (2016).; available 8/11/16 open access at:
http://jes.ecsdl.org/content/163/10/F1162.full.pdf

14. FangFang Li, Jason Lau, Stuart Licht,
**"Sungas instead of syngas: Efficient co-production of CO and $H_2$ from a single beam of sunlight,"**
***Advanced Science*, 2, 1500260** 1-5, DOI: 10.1002/advs.201500260 pub.online Oct. 1 (2015).

15. FangFang Li, Baohui Wang, Stuart Licht*,
"***Sustainable Electrochemical Synthesis of large grain or catalyst sized iron***,"
***Journal of Sustainable Metallurgy,*** 2, 405-415 (2016). Cover article, available open access
http://link.springer.com/article/10.1007/s40831-016-0062-8

16. FangFang Li, Shuzi Liu, Baochen Cui, Jason Lau, Jessica Stuart, Stuart Licht,
**"A one-pot synthesis of hydrogen and carbon fuels from water and carbon dioxide,"**
***Advanced Energy Materials, 7,*** 1401791-1-7 (2015). cover illustration for article at:
http://onlinelibrary.wiley.com/doi/10.1002/aenm.201570039/abstract

**Older publications particularly relevant to the C2CNT process**

17. Stuart Licht, **"**STEP (solar thermal electrochemical photo) generation of energetic molecules:
**"A solar chemical process to end anthropogenic global warming,"**
***Journal of Physical Chemistry, C*, *113***, 16283-16292 (2009).

18. Stuart Licht, Baohui Wang, Susanta Ghosh, Hina Ayub, Dianlu Jiang, Jason Ganley
**"A New Solar Carbon Capture Process: Solar Thermal Electrochemical Photo (STEP) Carbon Capture,"**
***Journal of Physical Chemistry Letters,*** 1, 2363-2368 (2010).

19. Stuart Licht, **"Efficient Solar-Driven Synthesis, Carbon Capture, and Desalinization, STEP: Solar Thermal Electrochemical Production of Fuels, Metals, Bleach**"
***Advanced Materials***, *47*, 5592-5612 (2011).

20. Stuart Licht, Baochen Cui, Jessica Stuart, Baohui Wang, Jason Lau,
"Molten Air Batteries - A new, highest energy class of rechargeable batteries**,"**
***Energy & Environmental Science***, **6**, 3646-3657 (2013).

21. Stuart Licht, Baochen Cui, Baohui Wang, F.-F. Li, Jason Lau, and Suzhi Liu,
"**Ammonia synthesis by $N_2$ and steam electrolysis in molten hydroxide suspensions of nanoscale $Fe_2O_3$**,"
***Science, 345***, 637-640 (2014).

## Additional (unpublished) graphs, data, materials and photos of Performance Data (delineating Sect. 4 of the submitted 25 page C2CNT written Submission)

Initial demonstrations of operation included in this description are the of carbon dioxide's extraordinarily rapid rate of carbon dioxide absorption from the gas phase into molten lithium carbonate and molten lithium carbonate mixtures was experimentally determined and documented. Even the lowest carbon dioxide concentrations studied (0.04% CO2 using conventional air) is sufficient to maintain and renew all molten lithium carbonate in an open air system during electrolyses conducted at constant current density of 0.1 A/$cm^2$. During the electrolysis, lithium oxide is co-generated at the cathode, which reacts with carbon dioxide continuously renewing the electrolyte. As demonstrated in the figure below, the rate of carbon dioxide absorption bubbled into even a small amount (50 g) of molten lithium carbon dioxide gas is not limited until the flow rates is well over 0.3 liters $CO_2$ per minute, and as expected (not shown) further increases with added lithium oxide concentration. The $CO_2$ cumulative absorbed on the vertical axis is limited to just below 100% due to the natural lithium oxide concentration which occurs on equilibrium with the lithium carbonate electrolyte (as we have measured in reference 11 on page 3 in the above section). After the measurements shown, it was noted that the calibration of the Omega CO2 controller calibration to measure and control CO2 flow rate had slipped downward, and the limiting Cumulative CO2 absorbed is higher than the indicated 92%. During the most rapid rate of C2CNT production tested of 1amp/$cm^2$, gas containing CO2 must be bubbled into the electrolyte, otherwise electrolyte would be consumed and the level of electrolyte visibly falls under those circumstances with bubbling a constant mass of electrolyte is maintained during the electrolysis.

**Experimental data**

| | |
|---|---|
| Electrolyte | 50 g Li2CO3 + 1 m Li2O |
| Crucible | Alumina, 100 mL |
| Temp, C | 770 |
| $CO_2$ flow rate, mL/min | 60-600 mL/min |
| CO2 inlet temperature | Room temperature (no pre-heating) |
| $CO_2$ inlet flow | Alumina tube: 1-bore (ID=3mm), sparger |
| measurements | Weight of crucible measured every 5 (or10 min), no electrolysis |

**Results**

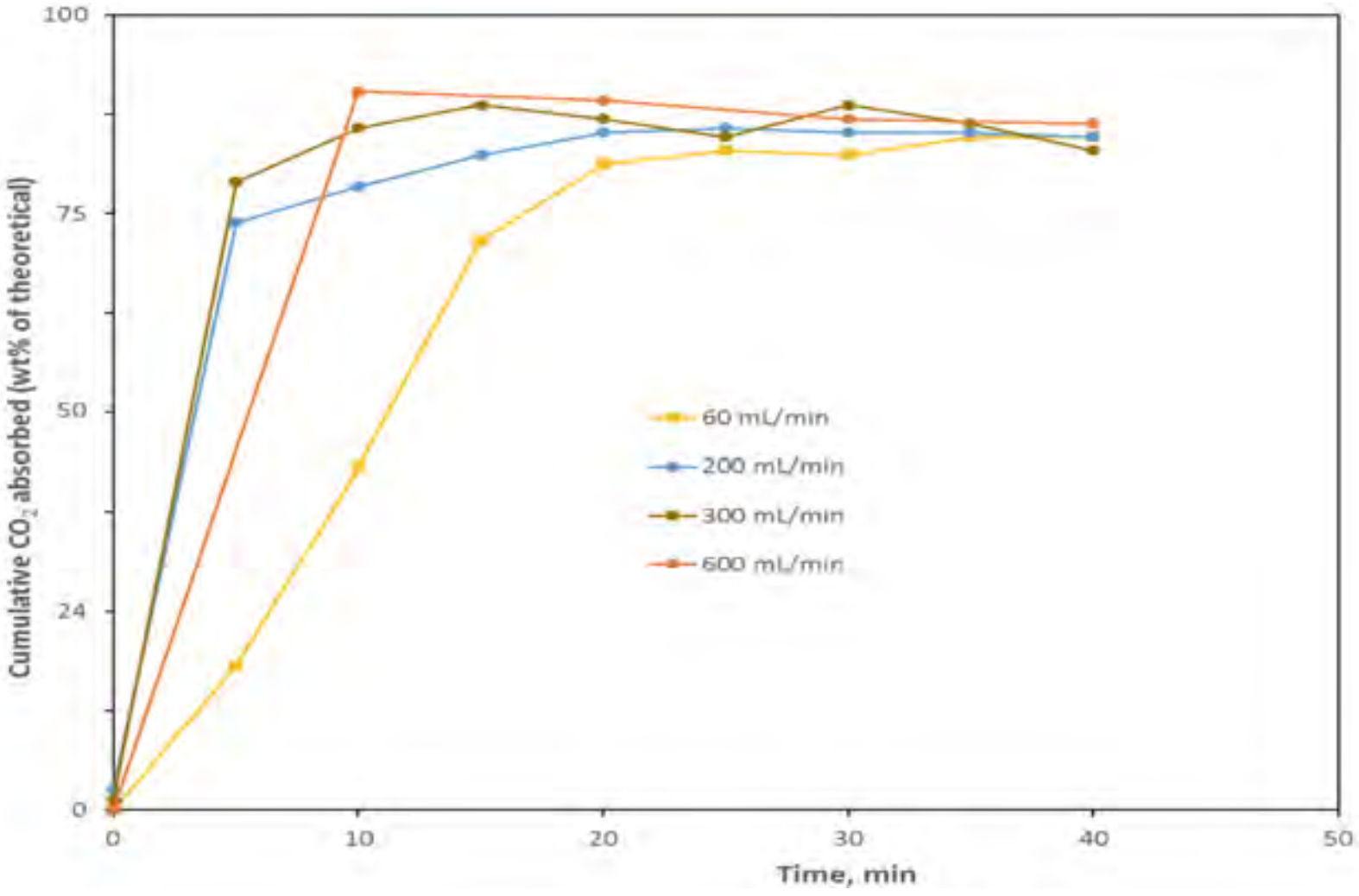

A large series of experiments conducted by C2CNT demonstrated that a wide range of interesting and useful carbon nanotubes can be uniformly formed by varying the molten carbonate composition, and the inexpensive metals used for the carbon nanotube generating cathode and the oxygen generating anode. <u>This is extensively, scientifically, photo documented in the references 1 through 11 on pages 2 and 3, and in the C2CNT scientific articles attached with the submission</u>.

A major breakthrough was reaching the critical thermal balance in which the C2CNT process requires no external heating even though it's a high temperature molten salt process. The lessons learned from the new technology of thermal energy storage using molten inorganic salts to store energy for concentrator solar powered towers where applied here. Molten salt thermal storage references include (i) Preito et al,, Solar Energy, 135, 518-526 (2016), (ii) Rodriguez et al., Appl. Thermal Eng., 63, 50-528 (2014), (iii) Peng et al., J. Thermal Therm. Anal. Calorim., 115, 1767-17777 (2014), Jonemann, Advanced Thermal Storage System, NREL subcontract report NREL?SR-5200-58595 (2013), (v) Sabharwall et al, Idaho National Lab., contract DE-AC07-05ID14517 (2010), (vi) Gabbrielli and Zampelli, J. Solar Energy Eng., 131, 041001 (2009),

The C2CNT process must first be preheated to the molten state. Subsequent to this the dissolution of carbon dioxide is high exothermic and generates heat. In addition, the electrolysis over-potential generates excess heat which grows proportionally with the electrical current density applied to the electrodes. In these July 2017 experiments, an in-house kiln was built (shown below using 9x4.4.x2.4 inch firebricks (purchased from BNZ, along with 24x9x4 inch firebricks for in-lab construction of the larger Genesis Device$^{TM}$) and the heating elements and control circuits from a commercial Pagargon Caldera kiln, and a custom thermal radiative shield was cut to be added as an intermediate kiln case from 0.034" thick mirror finish 304 stainless steel (purchased from onlinemetals.com) as seen in the photo to the right. One inch thick, of highly insulating rigid ceramic insulation was included barrier on all sides as a high temperature resistant, very low hear flow rate (K=0.28 at 800°C, purchased as Mcmaster.com product no. 6841K5 Extra-High Temperature Ceramic Insulation) thermal protective barrier, and is visible as the outer white edge, on the in construction kiln cover in addition to the grey furnace mortar, in the 2$^{nd}$ upper photo from the left below. Prior to the addition of the C2CNT CO2 to carbon nanotube electrolysis chamber, a 4$^{th}$ layer of thermal insulation (in addition to the firebrick, radiative barrier and ceramic insulation (was added as mineral insulation, yellow-green as seen in the photos below, and purchased as Mcmaster.com product no. 9328K43 2" thick Very-High Temperature to 65-C Mineral Wool Insulation Sheets)) to the kiln, and finally an outer coating of pink R-30 Home depot insulation was added as a final insulator and barrier to heat loss as seen in the photos below (prior to reinsertion of the electrolysis chamber(s) (both one or two electrolysis chambers were simultaneously included and tested in the insulated kiln). Then a final added R-30 exterior insulation is shown in the photos below (purchased from Home depot as conventional house insulation) prior to subsequent addition as the final outer layer of the kiln. Clearly the insulation can be further improved with thicker or better insulating materials, but the performance data below shows that the insulation in served its purpose fully. The figure below shows the kiln's ability to retain temperature as measured by thermocouple, subsequent to an initial heating to 800°C and then turning off the kiln, and it is seen that successive added thermal insulation barriers provided a high level of temperature stability.

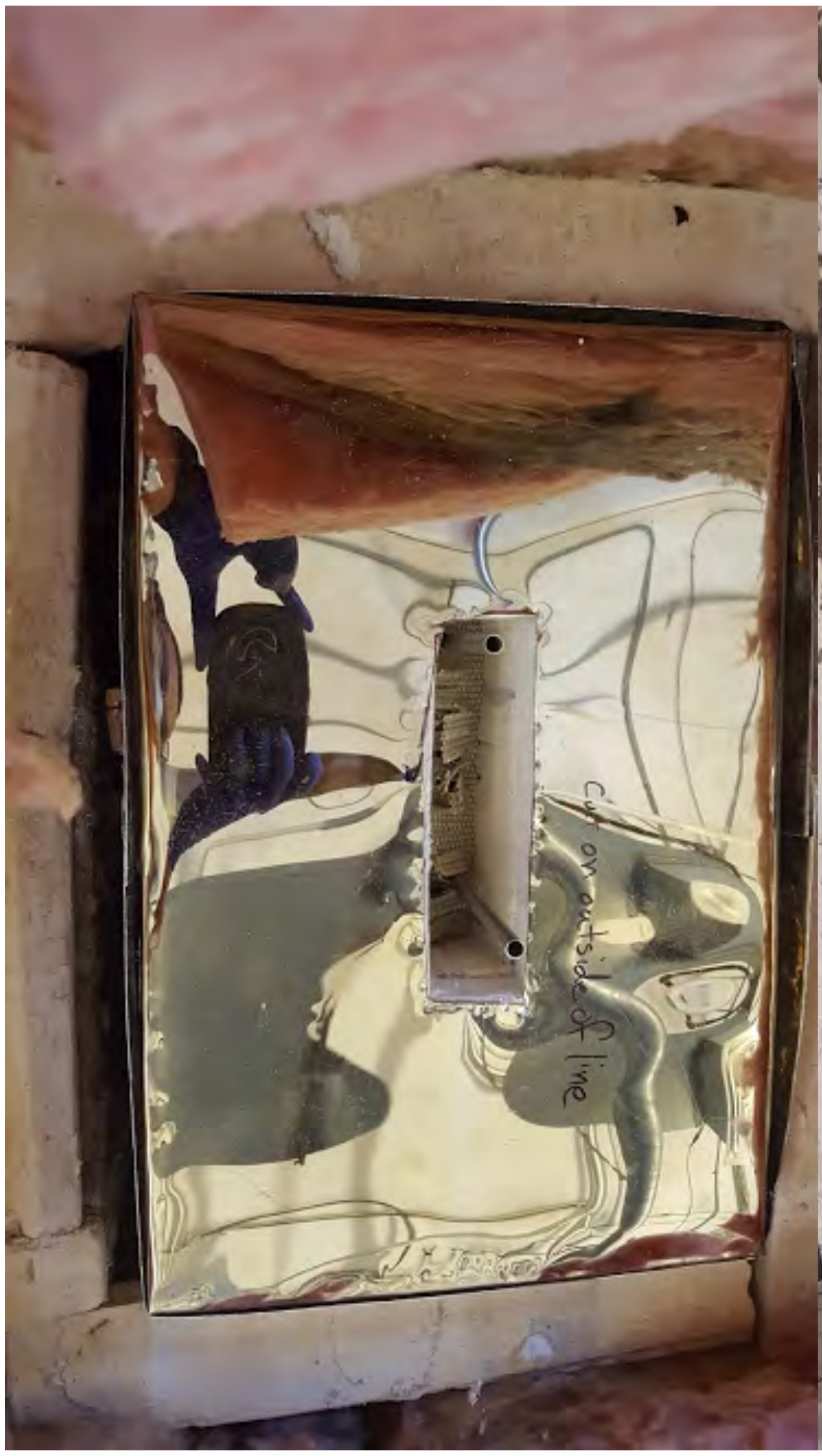
Cut on outside of line
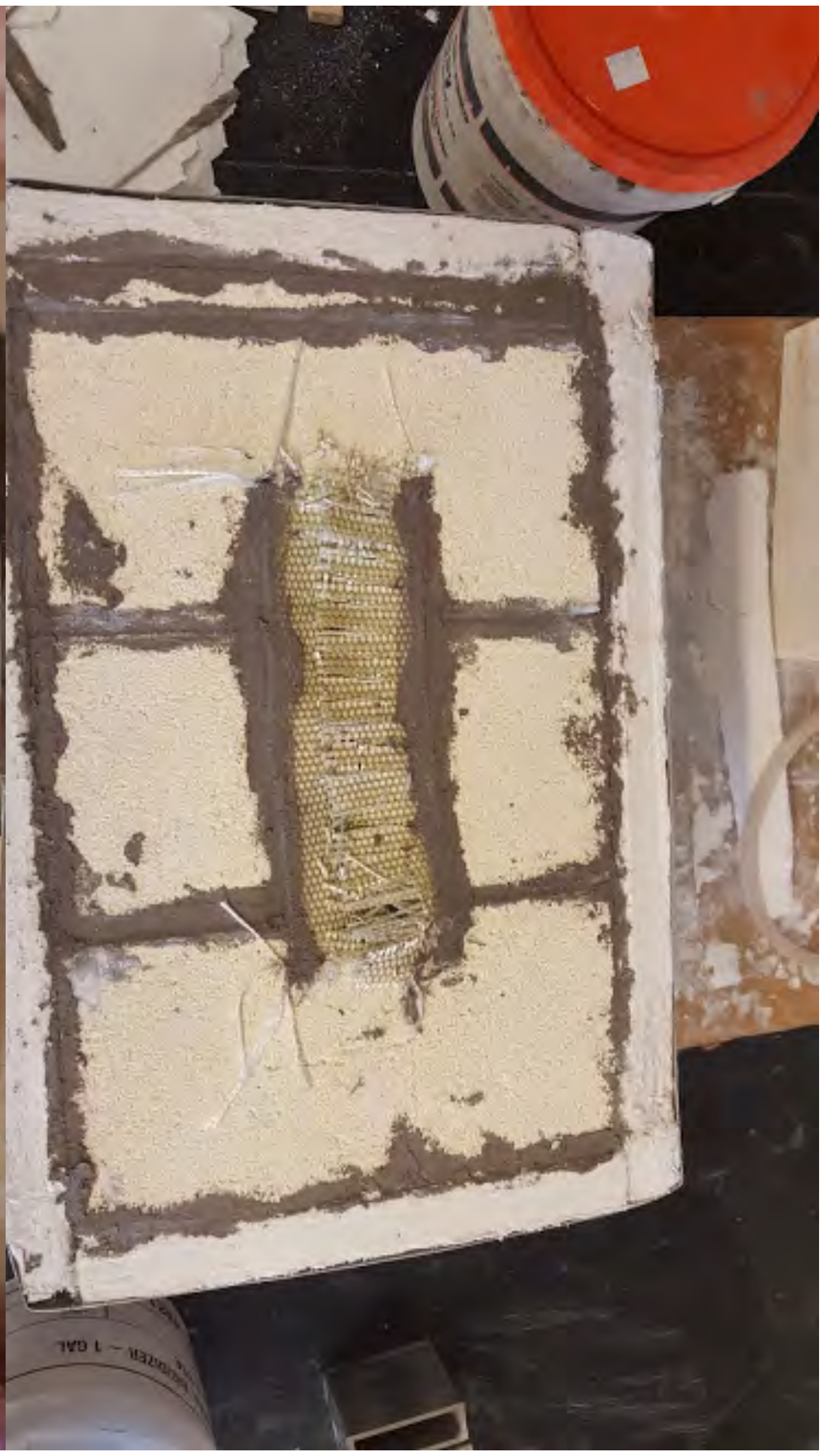
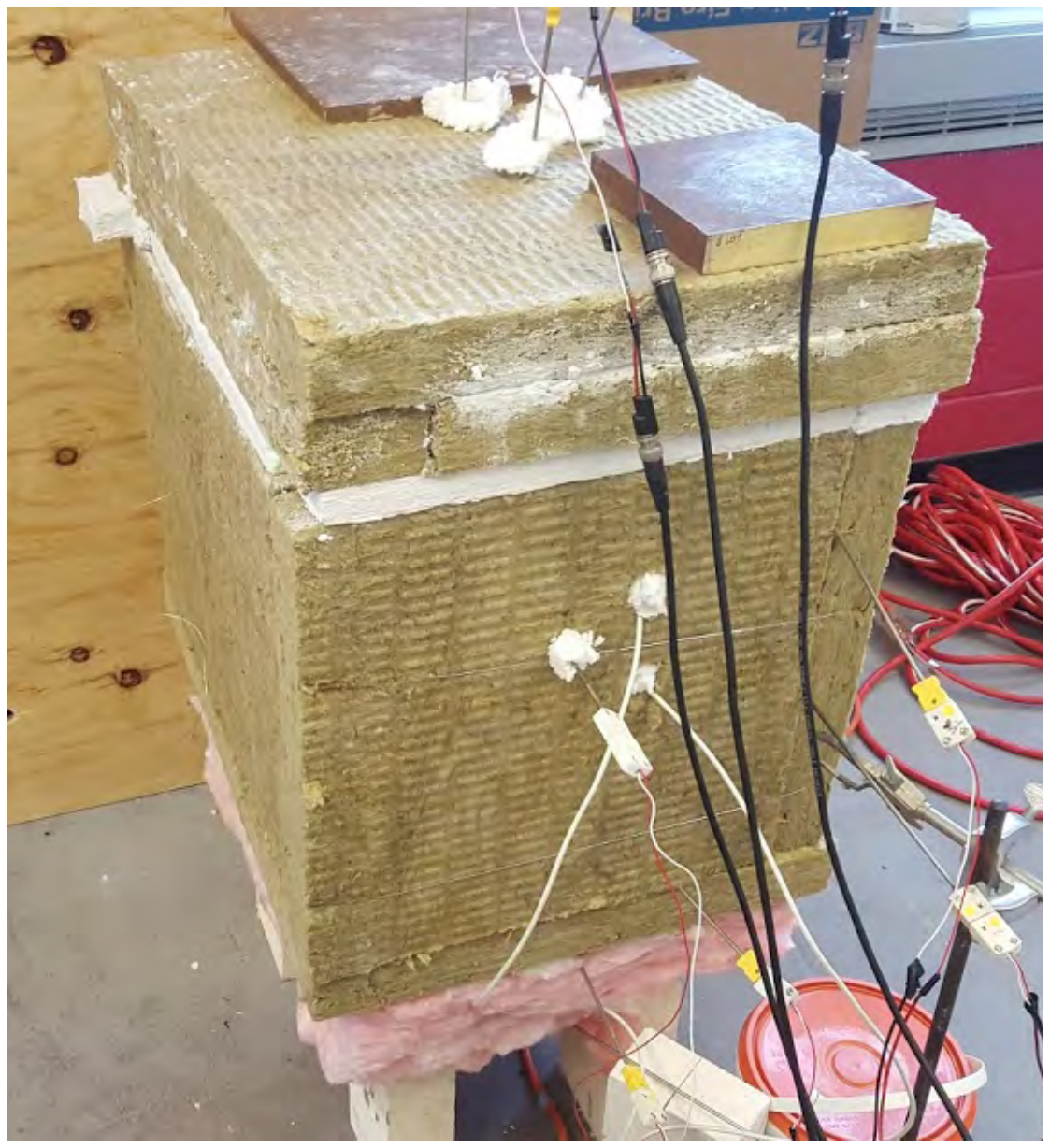
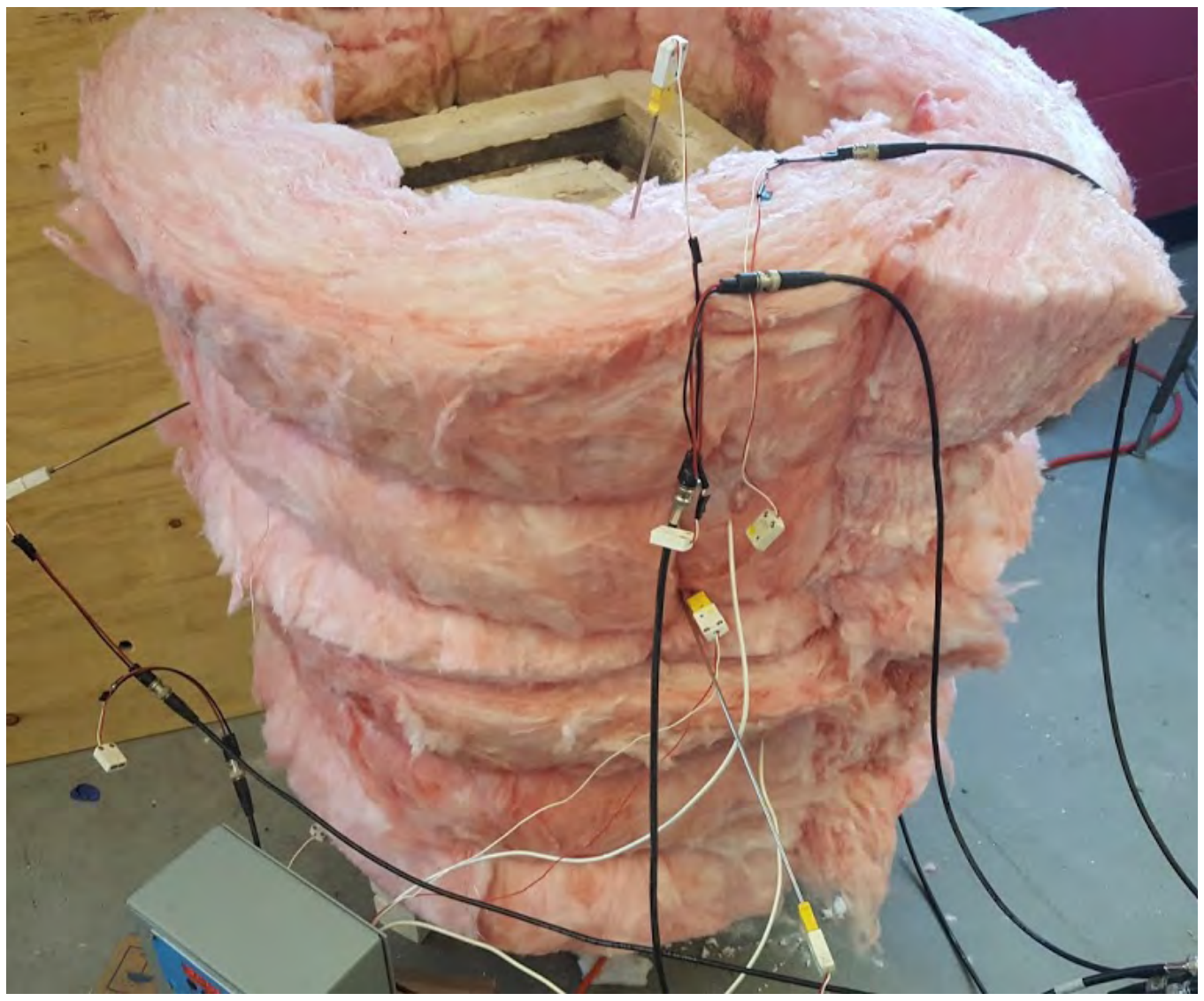

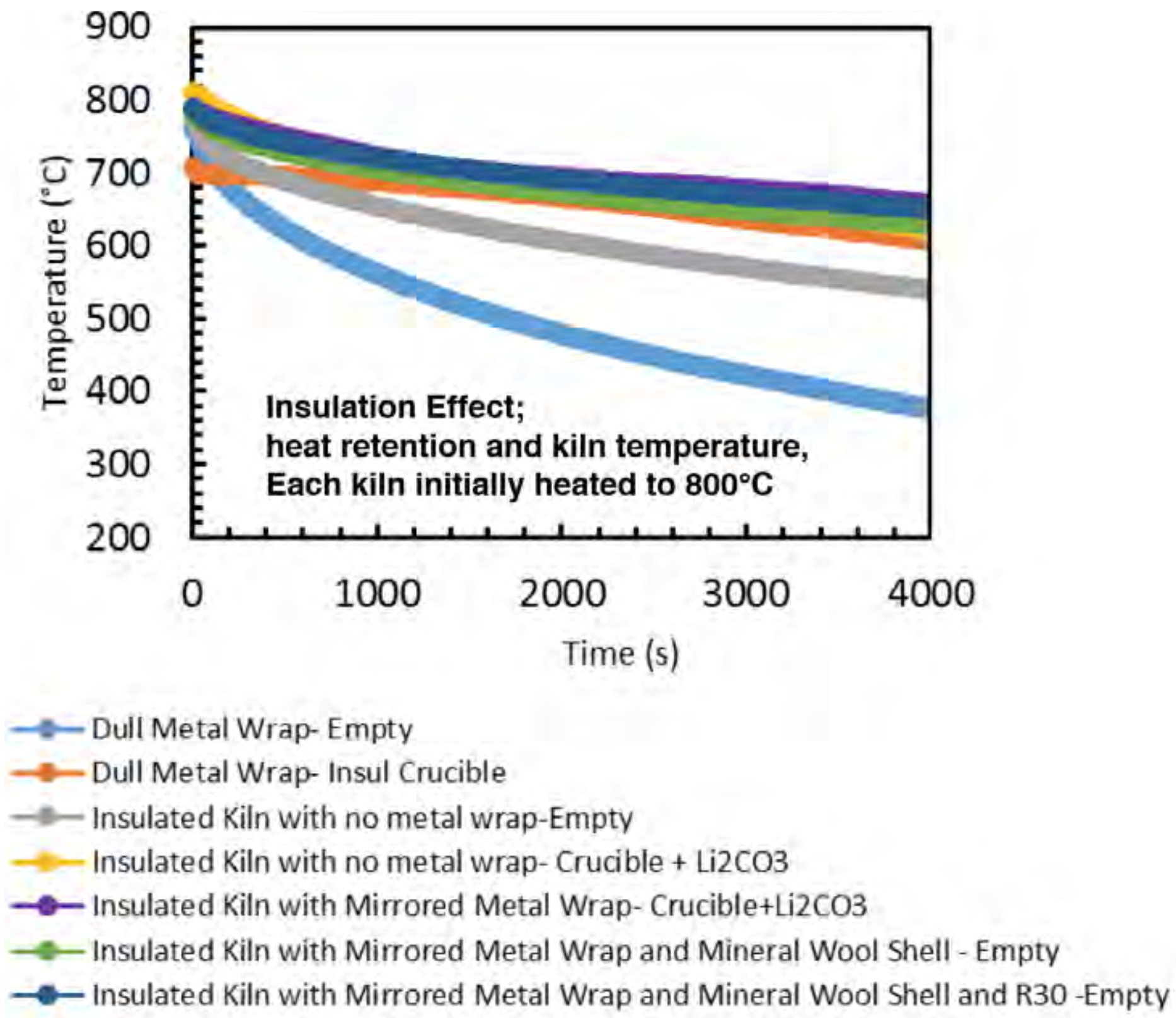

The custom built kiln with C2CNT electrolysis chamber was raised to 725°C (higher than the melting point of the lithium carbonate electrolyte), and then all kiln heating power turned off and the kiln was unplugged. At a 0.1 $cm^{-2}$ of constant electrolysis, the C2CNT independently maintained a constant temperature of 727°C. In accord with the expected exothermic nature of the $CO_2$ and $Li_2O$ reaction to constantly renew the electrolyte (and absorb the CO2), this temperature is observed increased to 737°C when $CO_2$ (unheated, pure) CO2 gas is bubbled in at a high precise rate comparable to the rate at which $CO_2$ is consumed by the electrolysis. This temperature increased to 787°C when the current density and (proportional $CO_2$ flow rate) is increased to 0.5 A cm-2, and decreased to 750° when the current density is decrease to 0.3 A $cm-^{-2}$ The figure below shows the performance at a continual of the constant electrolysis current density of 0.3A $cm^{-2}$ , and it is seen that this constant temperature of 750°C is maintained throughout the duration of the electrolysis.. The outlier temperatures in the middle of the experiment were due to a poor thermocouple connection which was remedied This experiment uses two internal C2CNT electrochemical cells connected in series and the electrolysis potential measured during this experiment is 2.2($\underline{+}$0.2)V throughout the experiment.

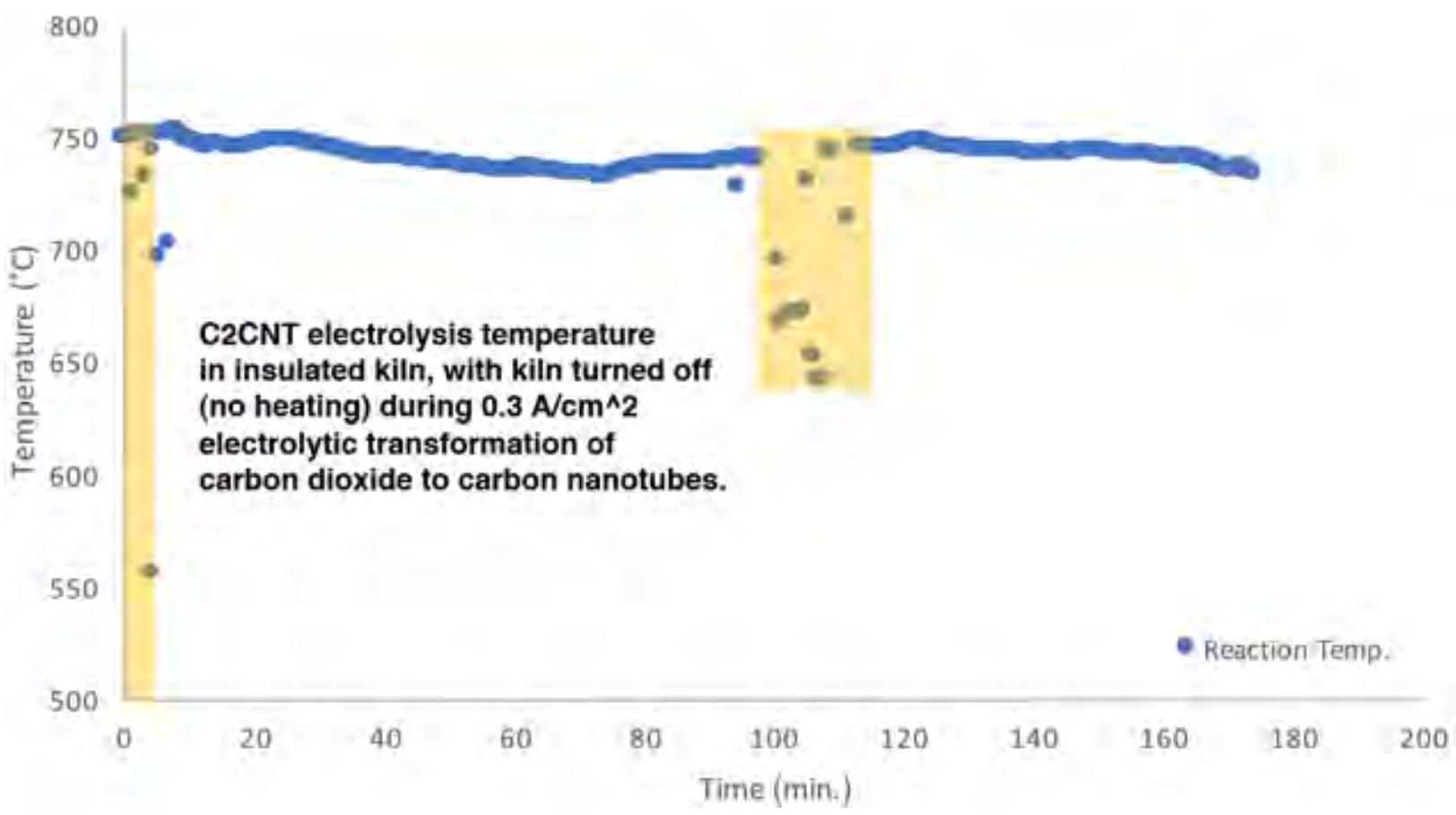

A variety of series and parallel electrode configurations were examined to optimize scaling up of the C2CNT process. Parallel electrodes allow us to effectively utilize a larger cumulative surface area electrode in a smaller volume. This minimized C2CNT surface area to volume ratio minimizes heat loss. Series electrode arrangements allows us to convert the maximum carbon dioxide within the voltage constraints of our purchased in-lab power supply ( a 10 V DC max, 4500 A max MagnaPower MS Series Programmable DC power Supply MSA10-4500/480, to run several C2CNT CO2 to CNT electrolysis cells within the 10 volt window of the power supply).

We observe that 0.135” stainless steel 304 cell makes an excellent C2CNT case material (we purchase this in 3 foot x 4 foot sections from Onlinemetals.com. We choice this 0.135” thickness only as it saves money It is the largest expensive per area, and least expensive of their thicker ss304 products). Stainless steel is substantially (an order of magnitude) less expensive than nickel. Nickel and many if its alloys are highly stable as alloys within the molten lithium carbonate and the basis of excellent C2CNT oxygen generating electrodes (anodes), but we observe that they are unstable as C2CNT case (external electrochemical body) materials, deteriorating after several days of use under hot atmospheric conditions.

A major portion involves using copper as an internal material in the metal electrode sandwich to provide sufficient conductivity and prevent voltage drop during the experiment. C2CNT experiments reported here are using inexpensive (non-noble, non-precious) metals and the type of metal has a large effect on the type of carbon nanotube grown from the carbon dioxide. For example, monel cathodes generate long carbon nanotubes and copper shorter cathodes, as well as some very small, nearly spherical carbon nanotubes (carbon nan-onions) as well as a small proportion of graphene sheets. Interestingly, while the shorter carbon nanotubes, the carbon nano-onions) and graphene have a higher current market price than the longer carbon nanotubes (graphene is priced at over one $million per ton), C2CNT feels the larger and more valuable future market is in the longer carbon nanotubes for larger applications such as metal replacements and textiles. Nevertheless, we are content that the XPrize collectively recognizes these various materials as a single carbon nanotube commodity with a current market value of $140,000 per ton. Metal resistivity is not an issue with smaller and intermediate electrodes, but was a potential major challenge with large electrodes. For the same current density (current per unit area), a small metal electrode will tend to have an insignificant voltage drop due to resistance loss. However, a long or thin electrode carrying a high current has a large voltage drop. Resistance drop this could have posed design challenges (energy losses) to optimal performance of large-scale C2CNT. Copper is highly conductive, and this challenge was solved by using pure or copper-clad anodes or cathodes. There are companies which produce to order custom clad electrodes for the aerospace industry, but this was beyond our price and time range. Instead, three alternatives were considered for C2CNT (i) the use of pure copper which became our choice for the cathode, but is insufficient (nickel and nichrome are better anode choices), (ii) a sheet metal of choice riveted or screwed onto an inner copper layer, or (iii) deposition (plating) of the metal of choice onto copper. We have purchased conventional, Chicago and blind rivets made from nickel, copper, brass and monel, and they function adequately to make a conductive contact from an outer thins sheet of brass, nickel or nichrome to an inner layer of copper, but are time consuming to install by hand. In addition to the use of simple 1/8” thick copper to use the for the cathode, we chose to electroplate nickel onto the 1/8” thick copper to use as effective anodes. The nickel plating was conducted in-house using a commercial nickel chloride aqueous solution, purchased from Epic Industrial, Inc., NJ, USA, with in-house added boric acid. All electrodes were cut from 3 foot by 8 foot sheets of 18” copper purchased from Storm Power Components, Decatur TN, SA.

C2CNT has been developed on a limited budget, which posed the challenge of generating the daily flue gas containing 200 kg of carbon dioxide in a cost effective manner. This challenge was solved in the following manner: by using air, rather than tanked nitrogen and oxygen, as the principal component to the flue gas. By its nature of having been used for the oxidation of fuels, post combustion flue gas contains less oxygen than air. We are grateful for the variance allowance to use air, and we additionally still use added tanked carbon dioxide (which contains 22.7 kg of $CO_2$ per laboratory tank) as a component of the flue gas. We use 7.3% CO2 in the round 2 test gas which fits within the allowed tolerances of both the Track A and Track B Carbon Xprize tolerances. The CO2 flow rate is carefully controlled and measured at 76 liter/minute (for the 200 kg daily CO2) by a 9/28/2017 calibrated Omega mass flow controller FMA5400/500 MASS FLOW CONTROLLER, which is built to control up to 131 liter/minute flow:

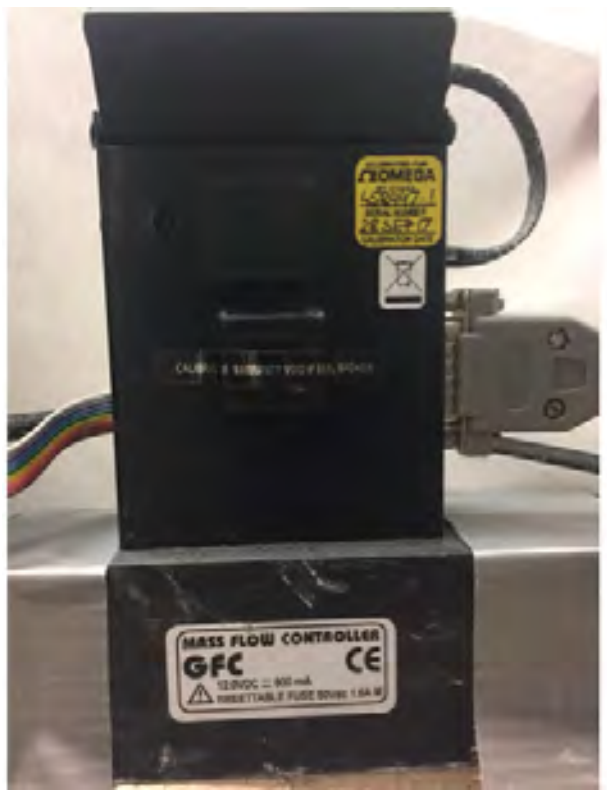

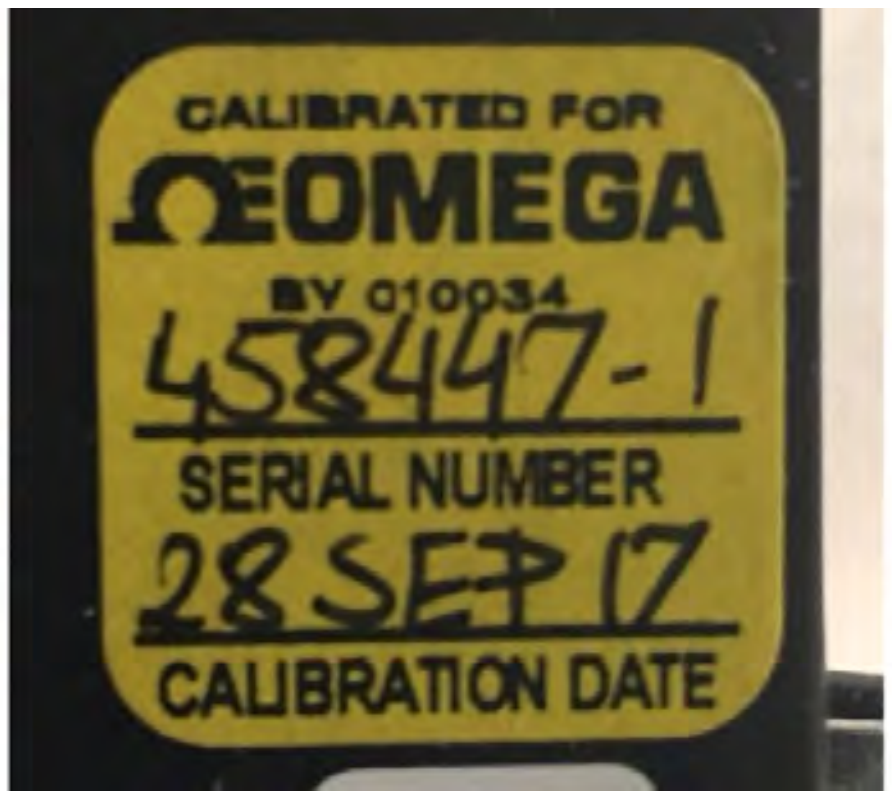

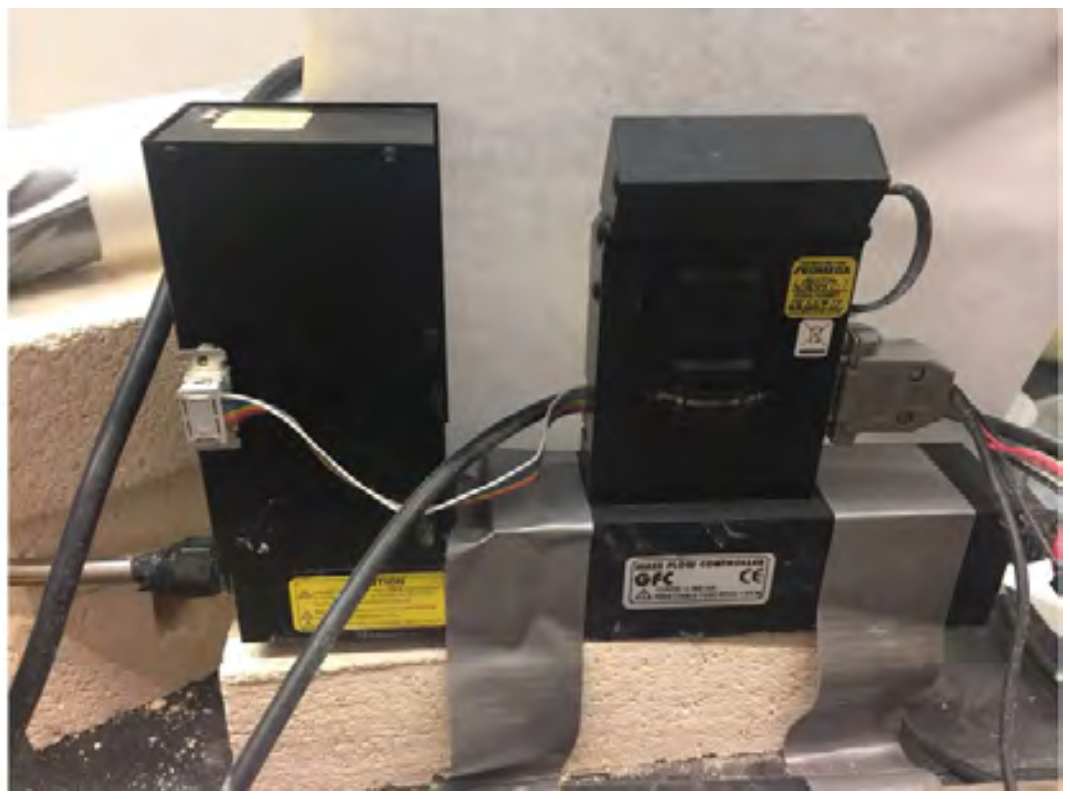

Air flow is provided with a VIVOSUN 4 Inch 190 CFM Inline Duct Fan with Variable Speed Controller (purchased for $59.95 from Amazon prime) providing a several fold excess of air (pictured below) and the air flow rate is redundantly monitored with an in-line Digi-Sense Hot Wire a Thermoanemometer with NIST Traceble Calibration and also with an in-line Petcaree Anemonter Sokos Digitial Wind Speed/Air Flow Thermometer.

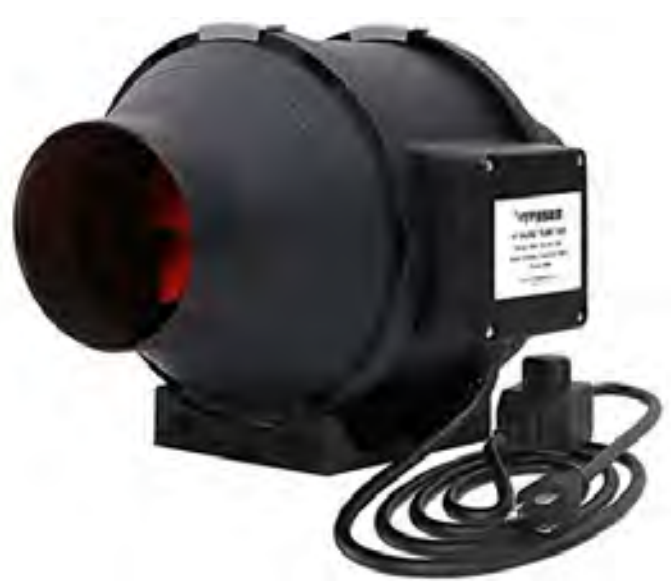

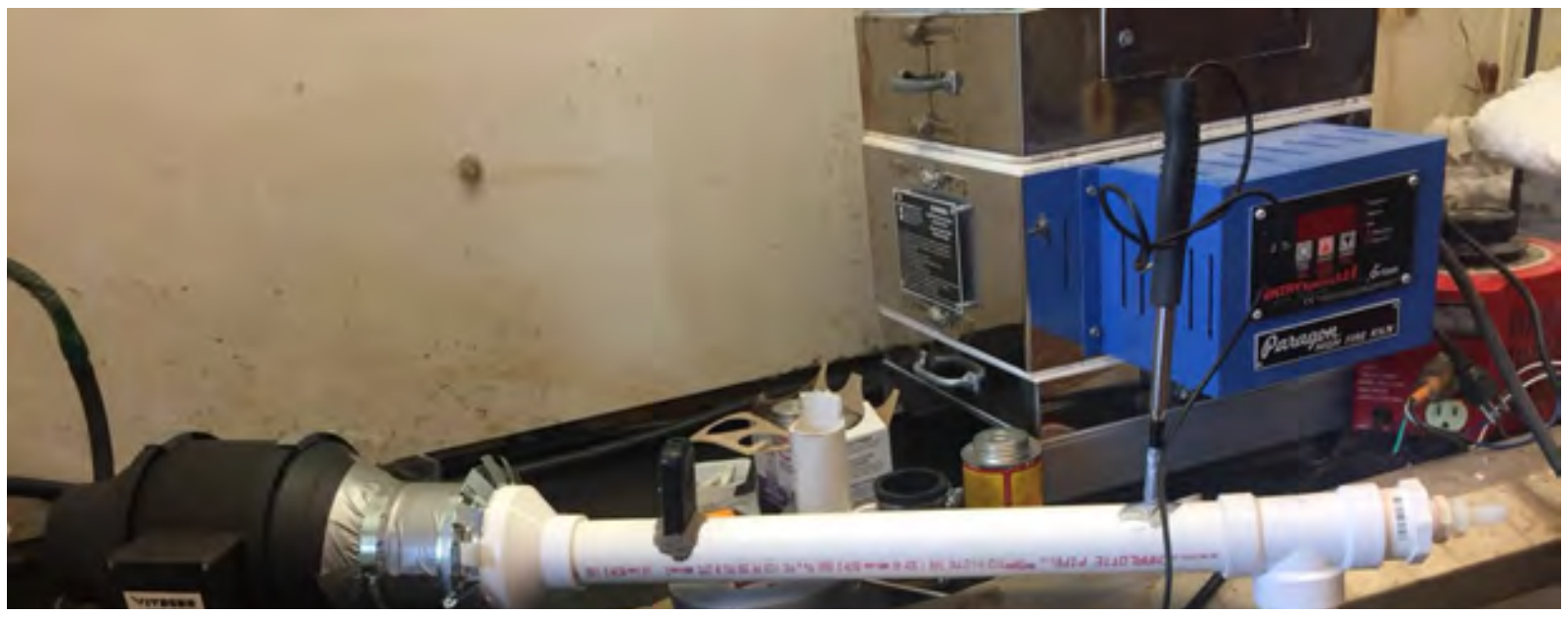

Particulate matter in the Track A gas flow has no observable detrimental effect on the C2CNT transformation of carbon dioxide to carbon nanotubes. Molten carbonates are highly insensitive to impurities. Particulate matter is either combusted, dissolved or (those with density greater than 1.8 g/cc) gradually sink to the bottom of the electrolysis chamber. We have accidently lost large and very small pieces of firebricks, insulation, mortar, broken ceramic tubes, plastics, and wires (with and without insulation), and tongs in the electrolyte without noticeable effect on the CNT synthesis. We have extensively documented in references 1 and 2 on page 2 that low levels of nitrogen, phosphate, sulfur, and boron oxide compounds have no noticeable detrimental effect on the product and that higher levels can imbue desired doped physical characteristics to the nanotubes (for example higher conductivity, catalytic activity and battery capacity).

Note, while C2CNT works proficiently with cold or hot gas, gas containing 7.3% CO2 and 92.7% air heated to 100°C (in accord with Track A temperature specifications) enters the C2CNT Genesis™ Device. Heating is accomplished by first passing the inlet gas through a simple heat exchange of a coiled tube entering and exiting a Caldera Paragon kiln, although a simple propane heater or simple coiled electric heater has also been used instead. NOx, $SO_2$ and CO in the correct Track A ppm proportions are also simply continuously added through the Duct Fan inlet prior to entering the C2CNT Genesis™ Device. To save money on lecture bottles and regulators for addition of these pollutants, they are generated in lab. NOx consists of NO and/or $NO_2$. NOx is generated by the reaction of copper metal with nitric acid; the rate is controlled by acid strength and relative thickness of the copper (). More NO is produced at lower nitric acid concentrations (4 molar NO generation as photo documented on the left side), while pure brown $NO_2$ is formed in concentrated nitric acid (as photo documented on the 2nd to left photo below). The 4 molar nitric acid gradually turns from colorless to blue as the $Cu^{2+}$ enters the solution. We chose to use this concentration of nitric to produce the NOx solely for aesthetic purposes and the darker $NO_2$ is equally easy to produce. Similarly, $SO_2$ is produced by the direct reaction of sulfur powder with sulfuric acid. We have a great deal of expertise in the controlled production of CO. We again use, but on a very small scale with a temperature controlled band heater, 100 ml alumina crucible, $Li_2CO_3$ electrolyte, coiled Ni wire $O_2$ anode, and coiled steel cathode, insulation and ceramic tube to direct the output direction of the CO directly into the inlet of the Duct Fan) the electrolysis of $CO_2$, but at 900°C, rather than in the 700°C temperature range. Although the CO product is hot, it is produced as such low required quantities that it does not impact the duct fan. The pair of coiled electrodes below are an extra set for diagrammatic purposes (there is an operating set within the molten electrolyte) As we have documented, at this higher temperature the product is pure CO (and $1/2O_2$), rather than pure carbon nanotubes (and $O_2$). This is a two electron, rather than four electron reduction of carbon dioxide and the rate of CO production is strictly controlled by the set rate of constant applied current (as delineated in references 14 and 18 on page 3 herein).

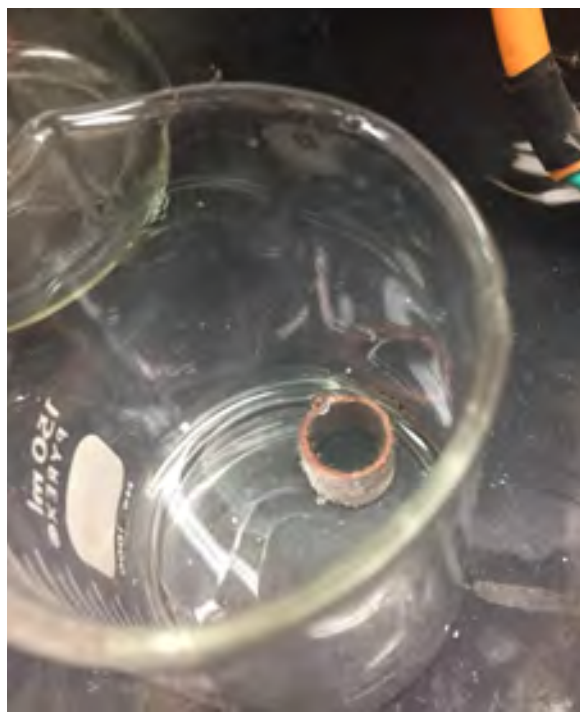

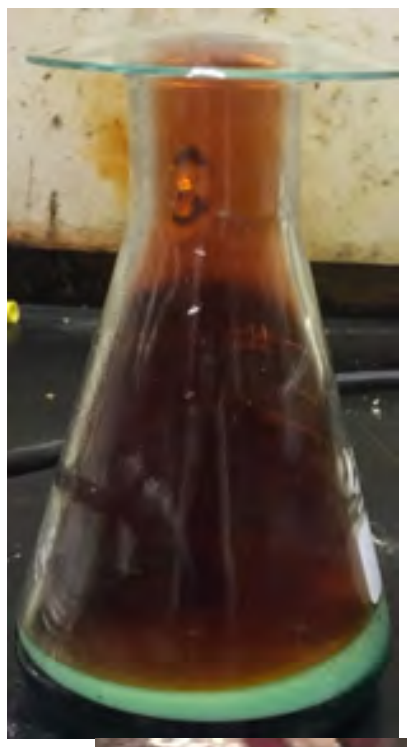

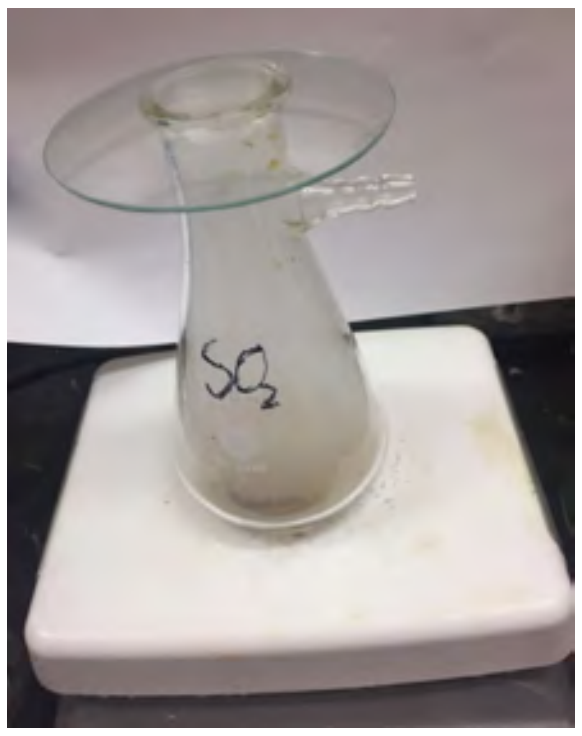

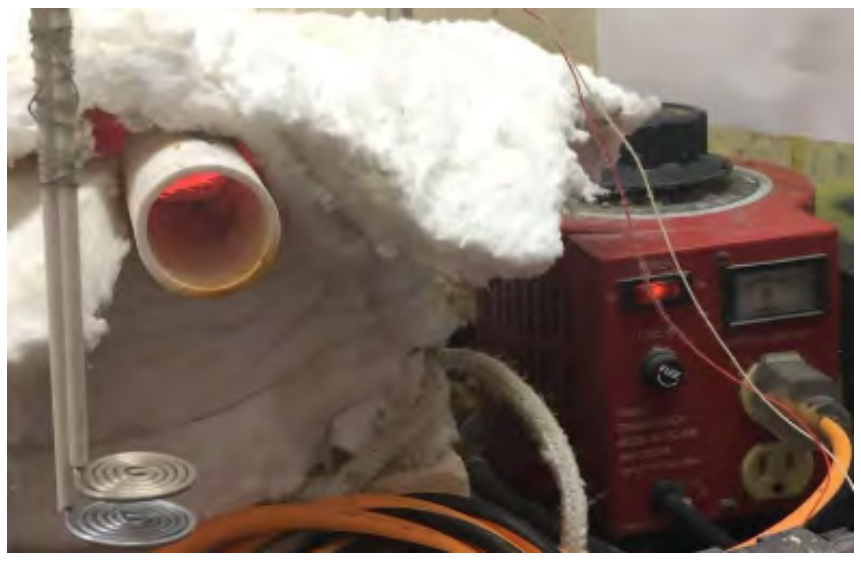

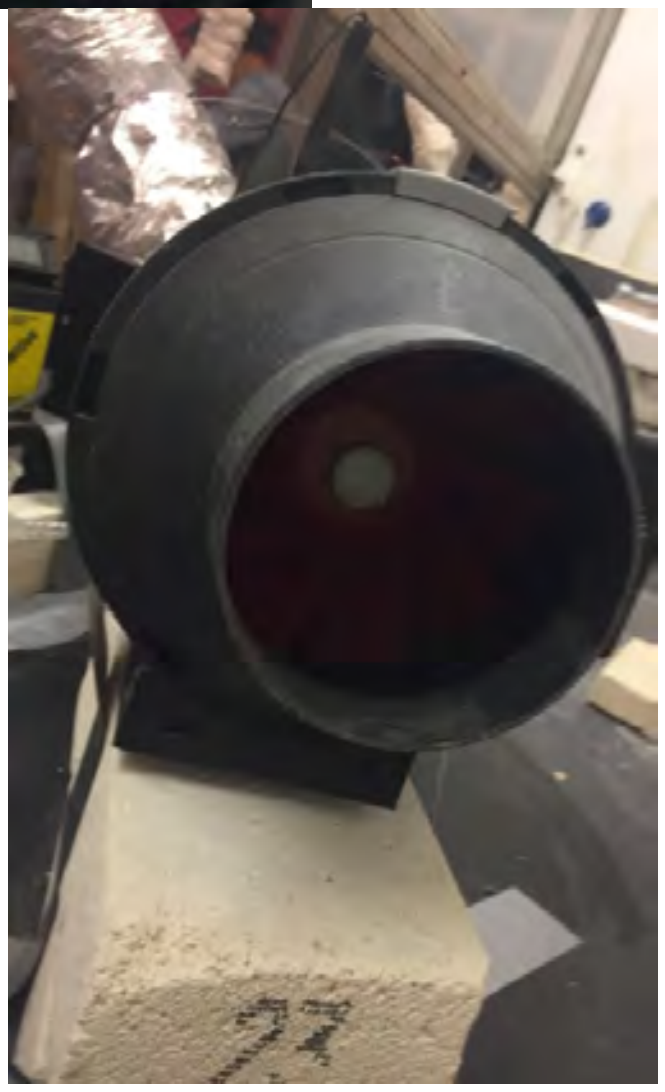

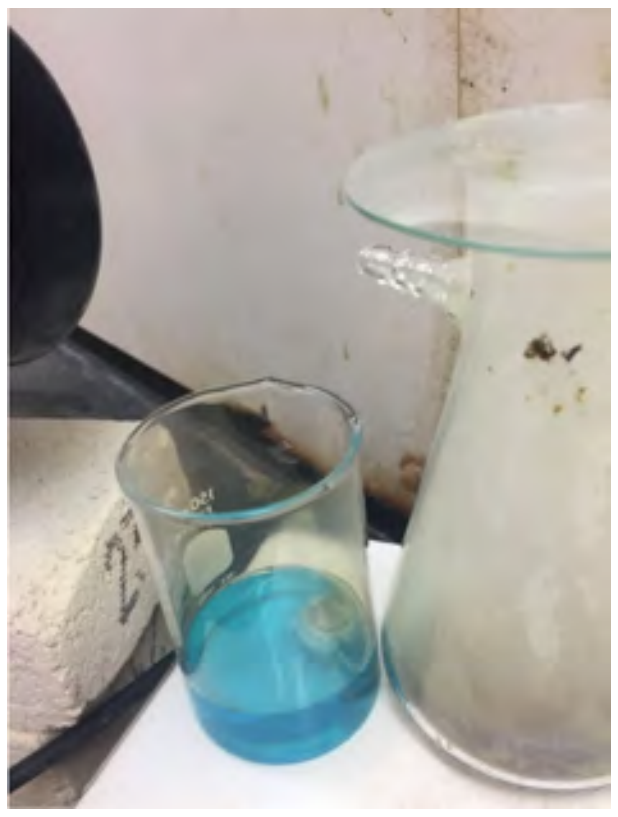

The electrodes are closely spaced in the 200kg CO2 daily Genesis Device™ and a means was devised to evenly disperse the inlet gas among the C2CNT electrolysis electrodes. It is shown below on the small scale; several layers of conventional large mesh copper screen are placed at the bottom of each electrochemical cell along with a tube to direct the gas among these layers. The (solid, powder) lithium carbonate electrolyte is placed in the cell and melted, the cathode (in this case monel, the thin sheet on the left side in the photo) and the air anode in this case, a thick nickel on the right side of the photo) are placed in pure alumina ceramic tubes to insulate them and apply pressure to the screen. Small gas bubbles disperse evenly through the electrolyte.

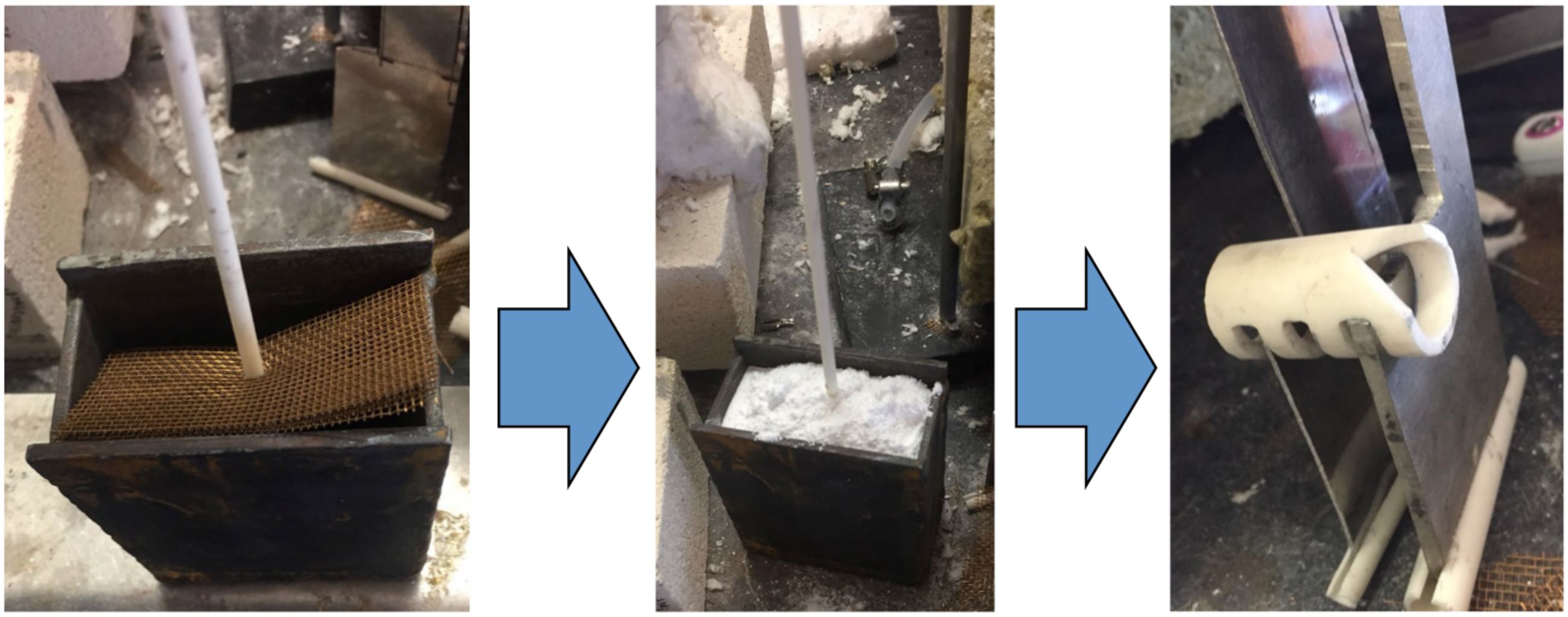

In the above test case conducted on Sept 22, 2017, , the inlet gas used was 5% $CO_2$ ) controlled by a smaller, similar OMEGA mass flow controller, mixed with 95% pure nitrogen (we had a nitrogen tank next to this electrolysis unit). The electrolysis was conducted at a constant 0.2 A $cm^{-2}$ current density, and as shown the product subsequent to electrolysis is pure carbon nanotubes

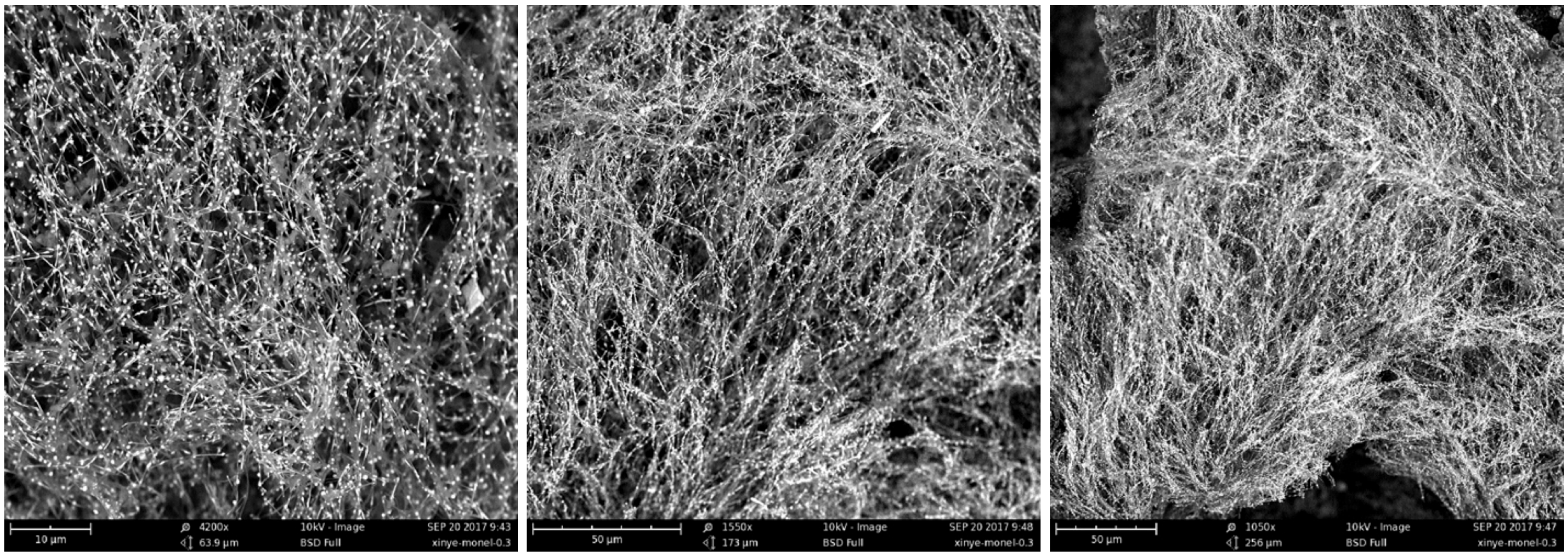

The described performance elements have been combined to build the 200kg of $CO_2$ daily transformed to carbon nanotubes C2CNTs Genesis device™. This Genesis Device™ is photo documented below, first in 2D-and 3-D representations and then in the various stages of the completed device. The representations include empty steel cells, cells with inserted anodes and cathodes, two cells placed in the C2CNT kiln, cells in the kiln with bus bars for intra cell parallel electrical connections, inter cell series electrical connections, tall bus bar connections to the MagnaPower supply, and electrical interconnects situated above the cover (outside the Genesis kiln).

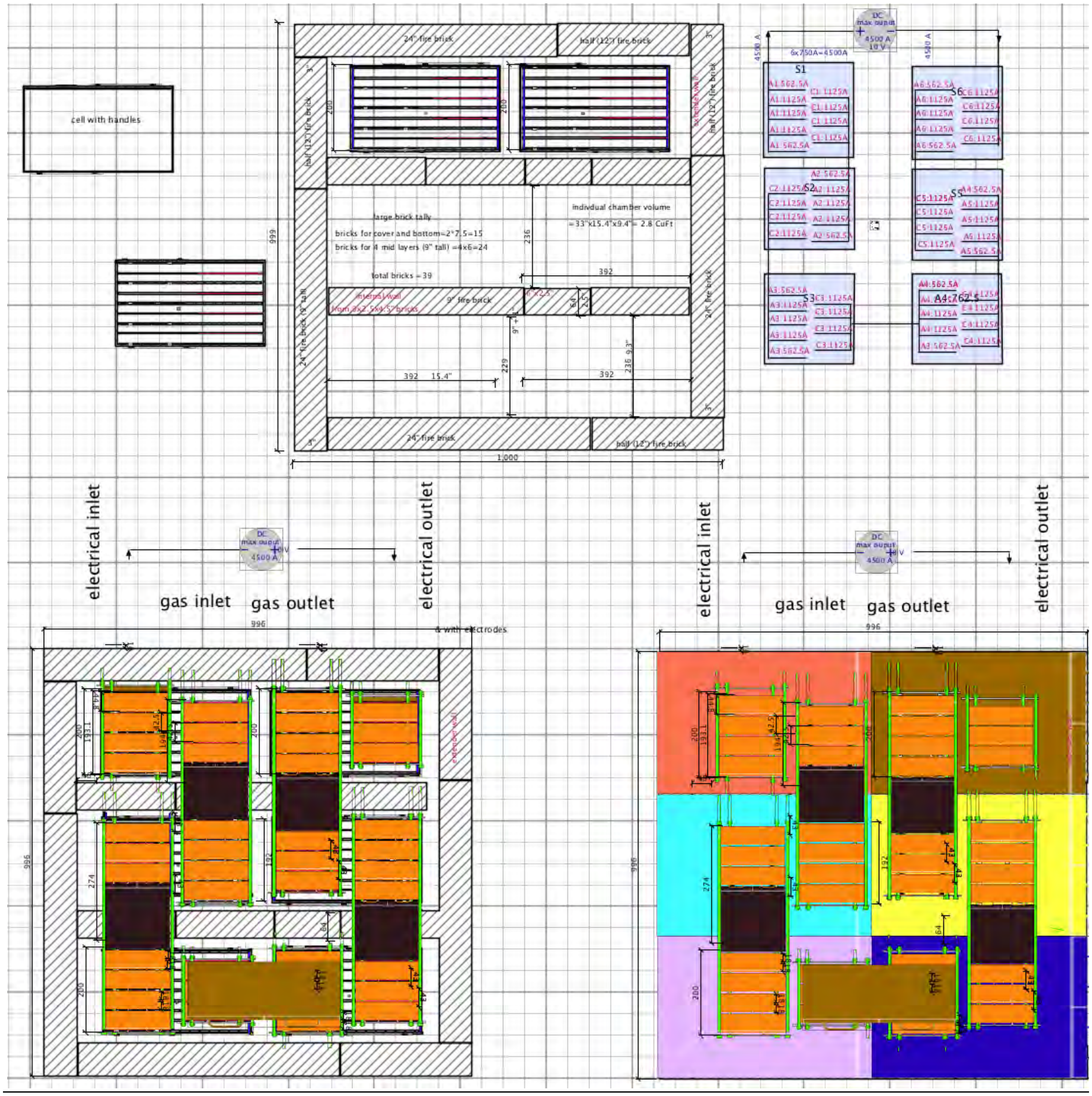

Three dimensional structure of the C2CNT Genesis Device™ represented in two dimensions in the previous figure.

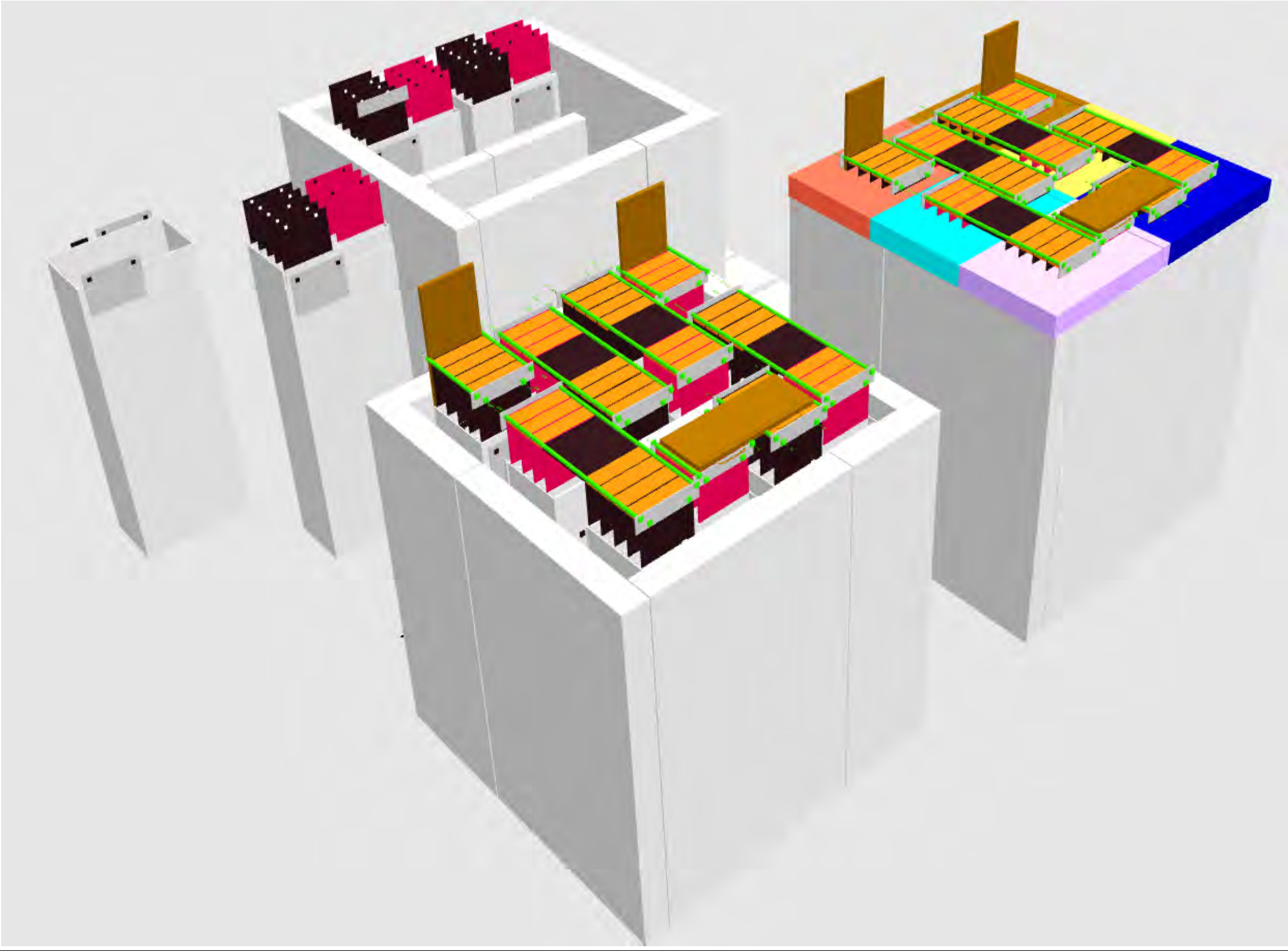

.

Left: Anodes post plating nickel onto 1/8" copper substrates. Right: Stainless steel 304 cases without components.

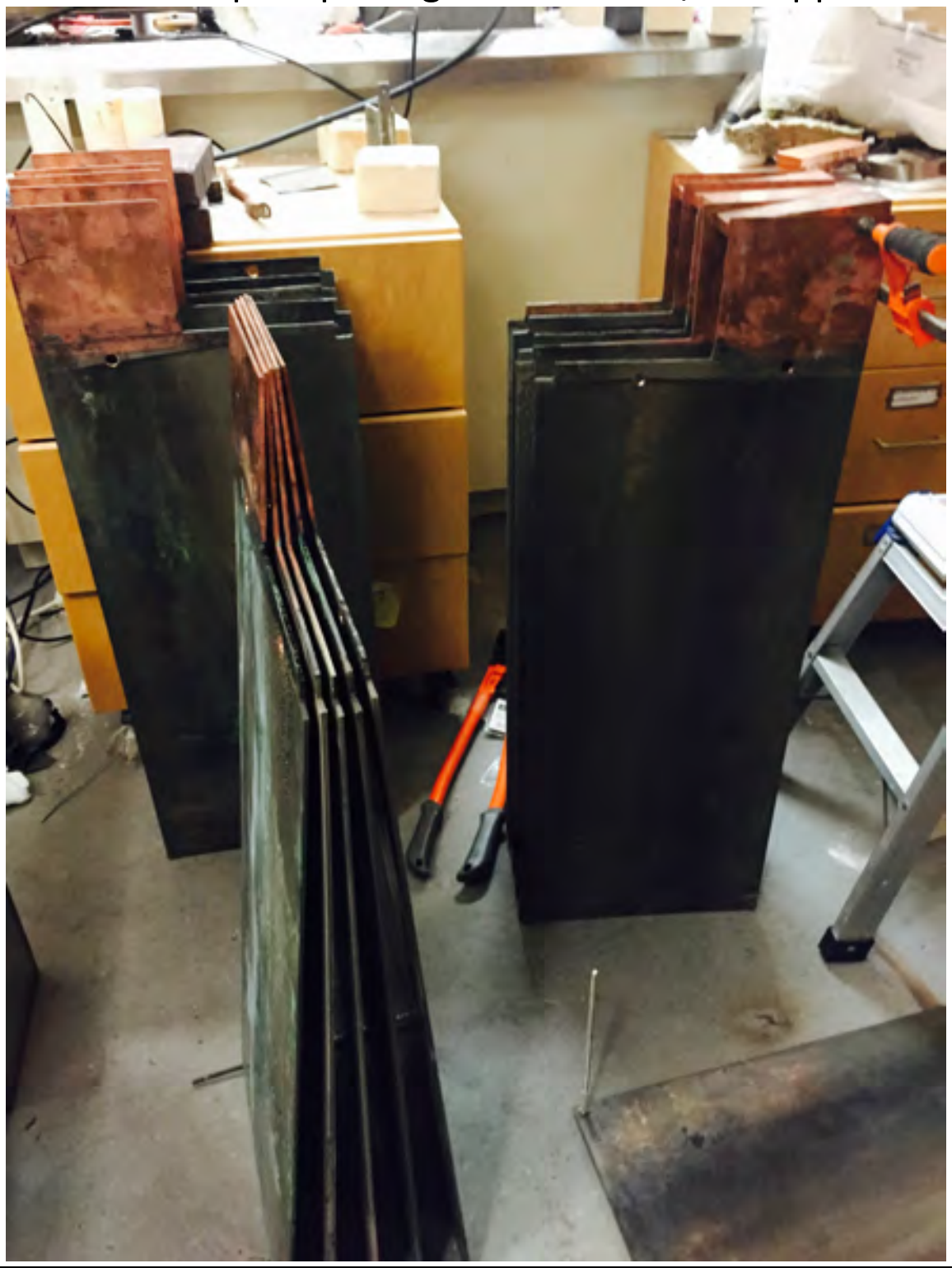

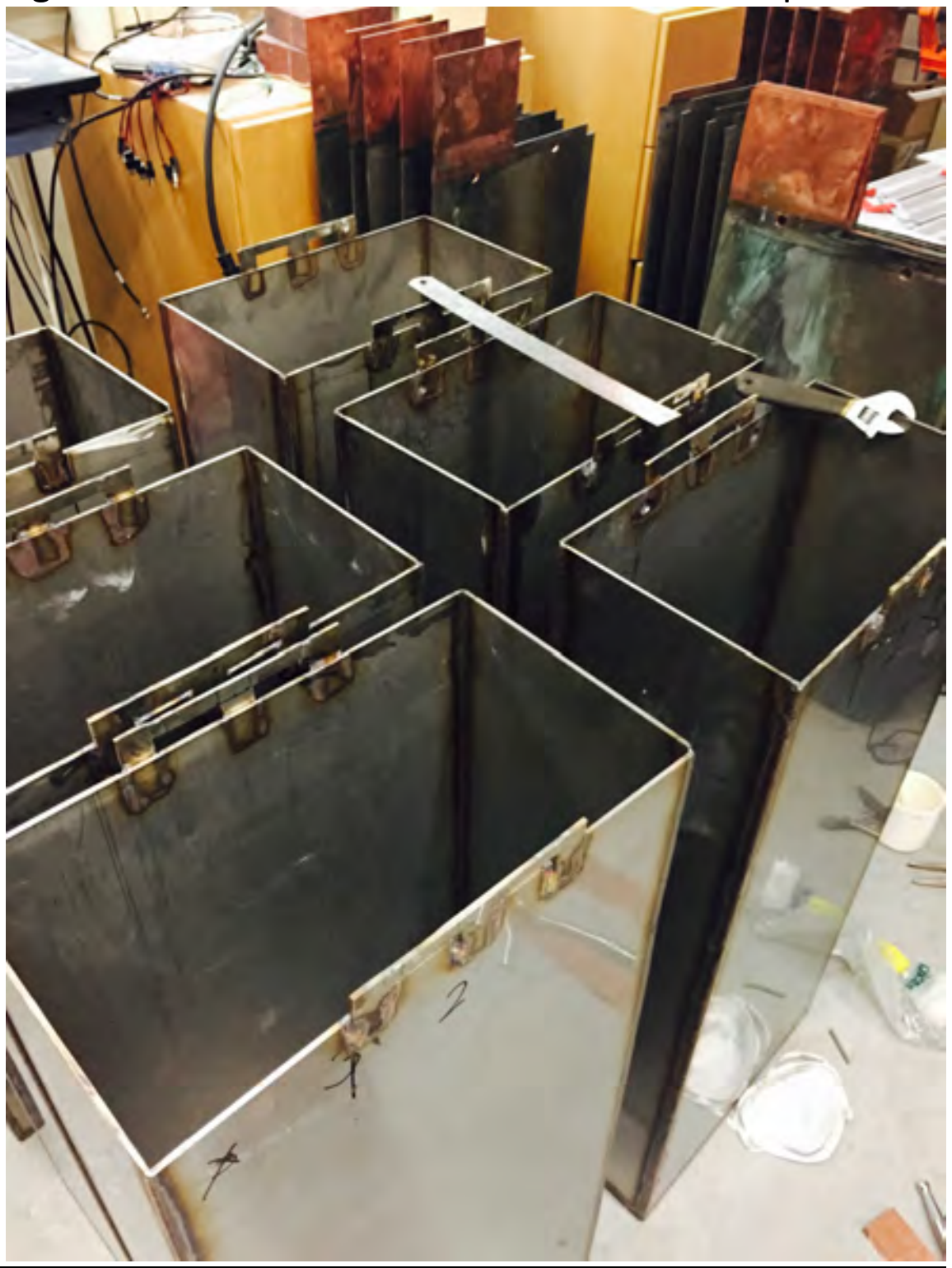

Left: The bare firebrick structure of the Genesis Device™.

Right: Addition of the mintherm insulation (not shown) and the shown mirror finish stainless steel 304 radiative shield casing.

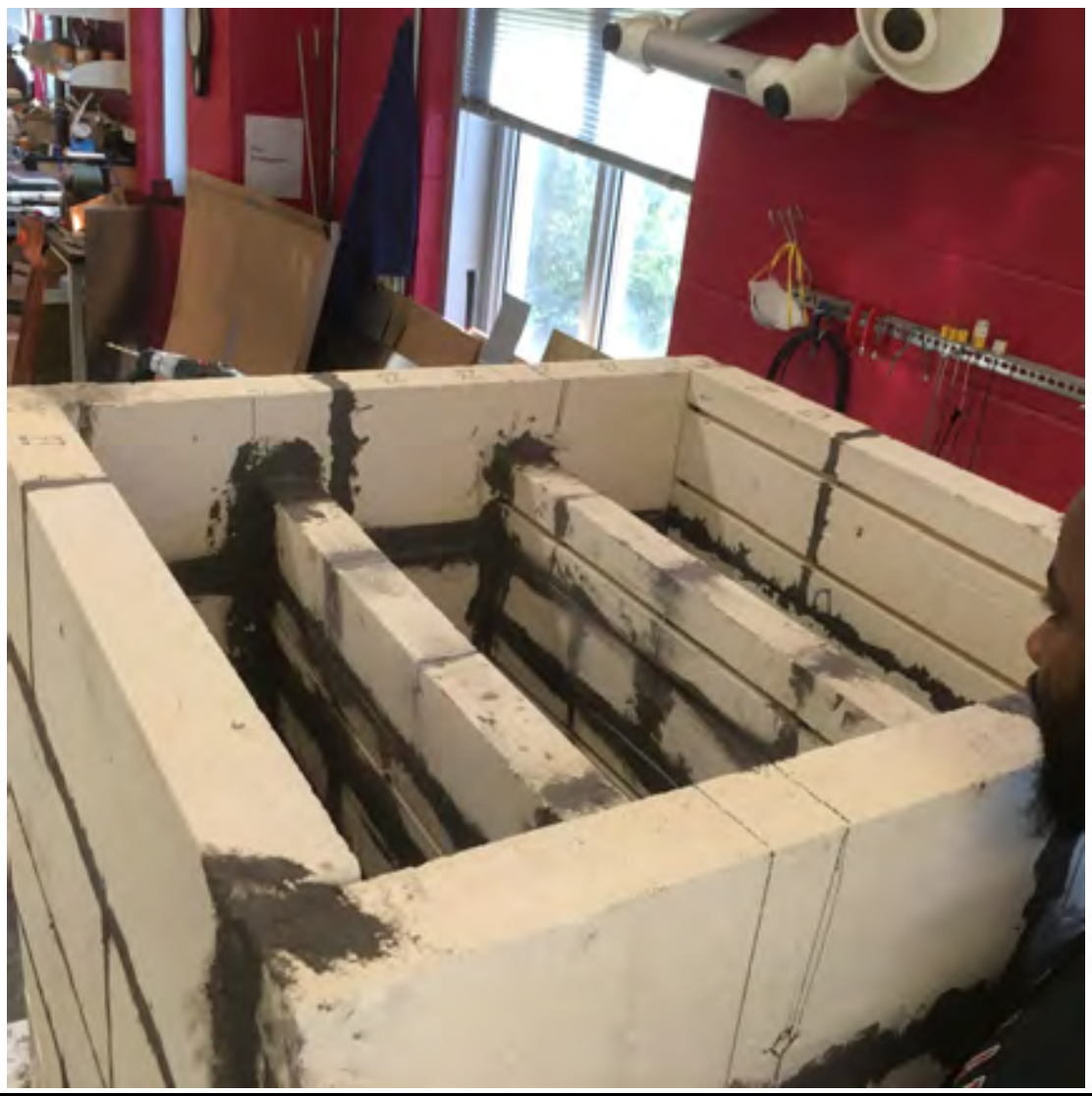

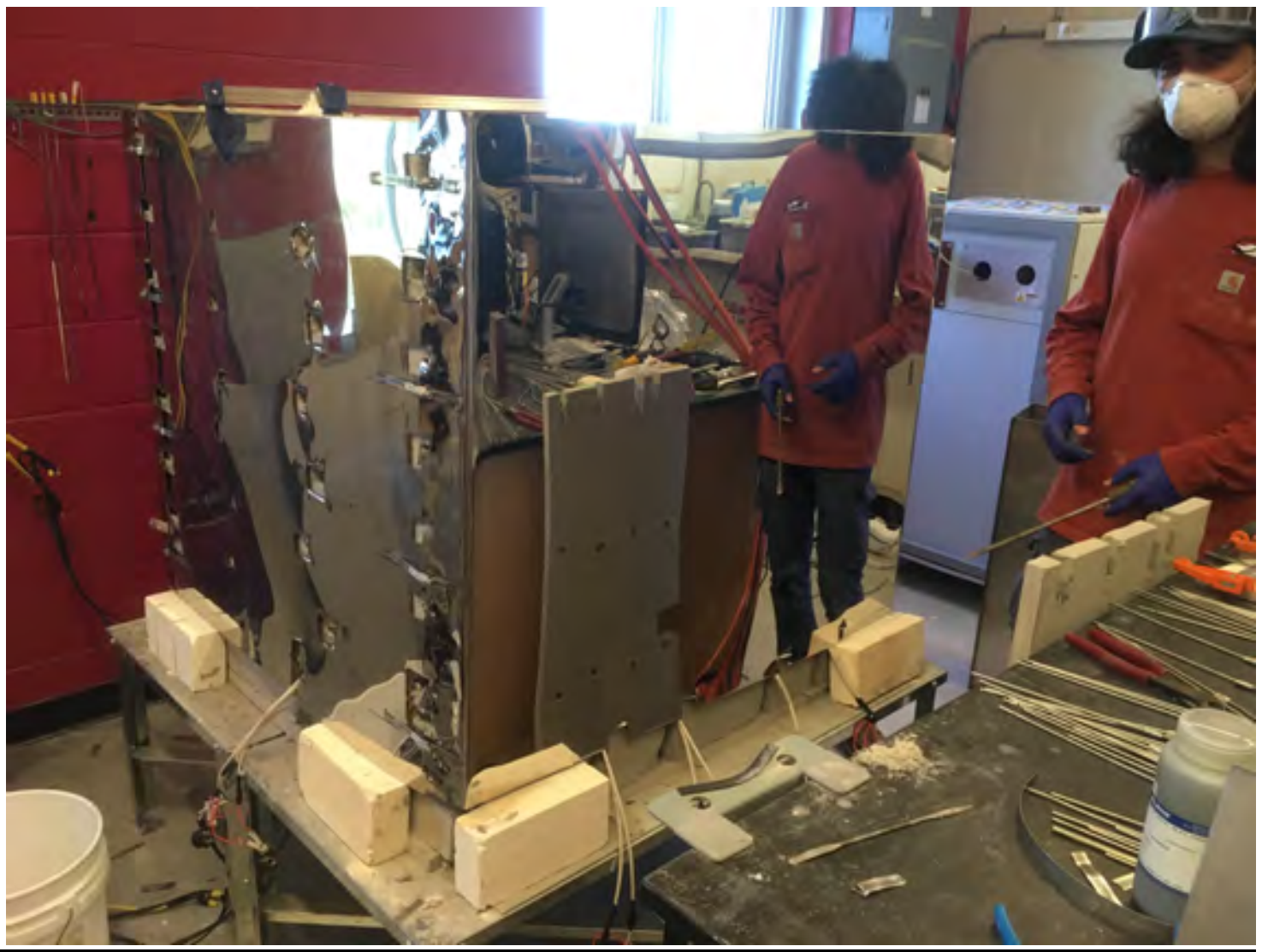

Left and right: Final adjustment and electrode check before addition of the lithium carbonate electrolyte.

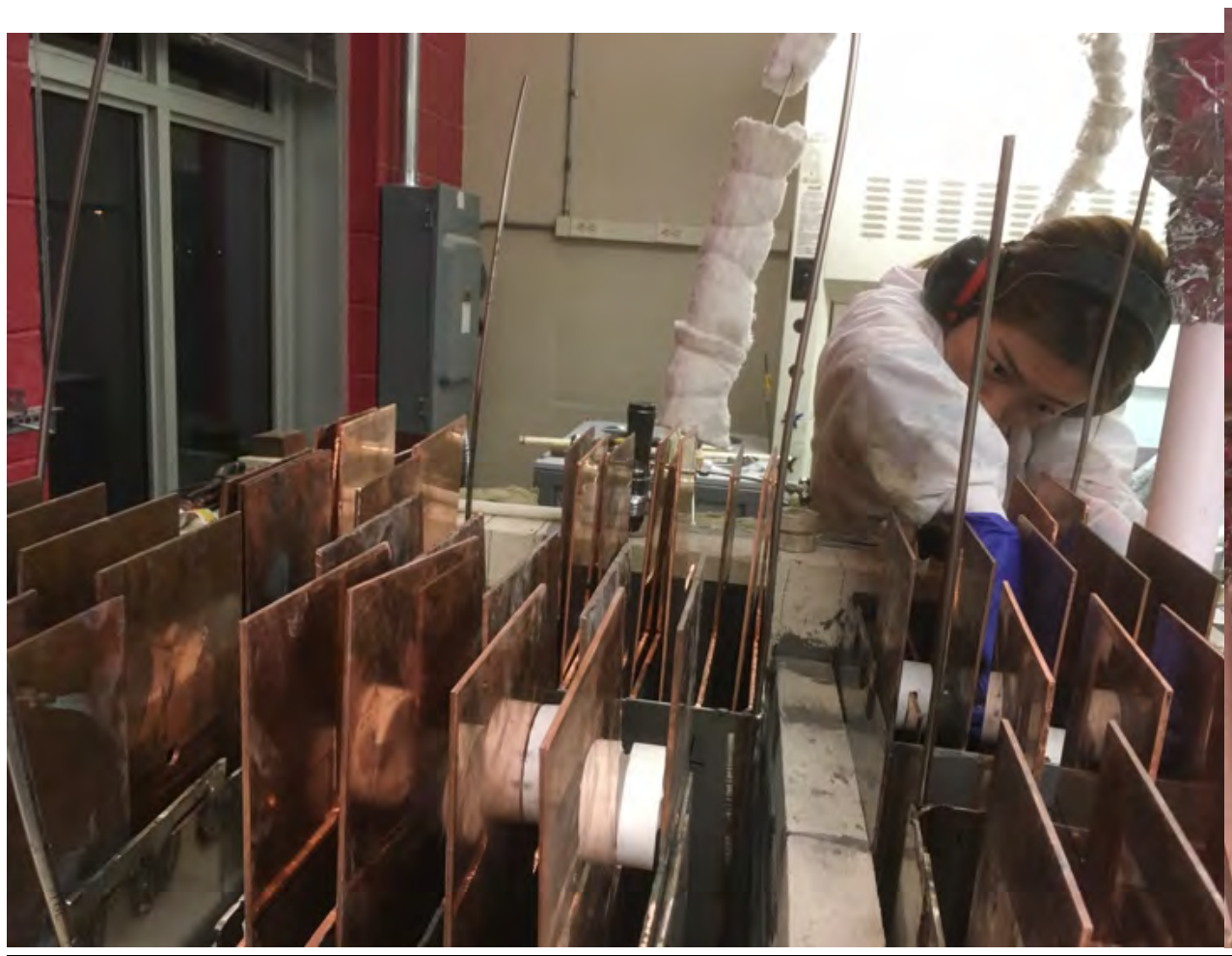

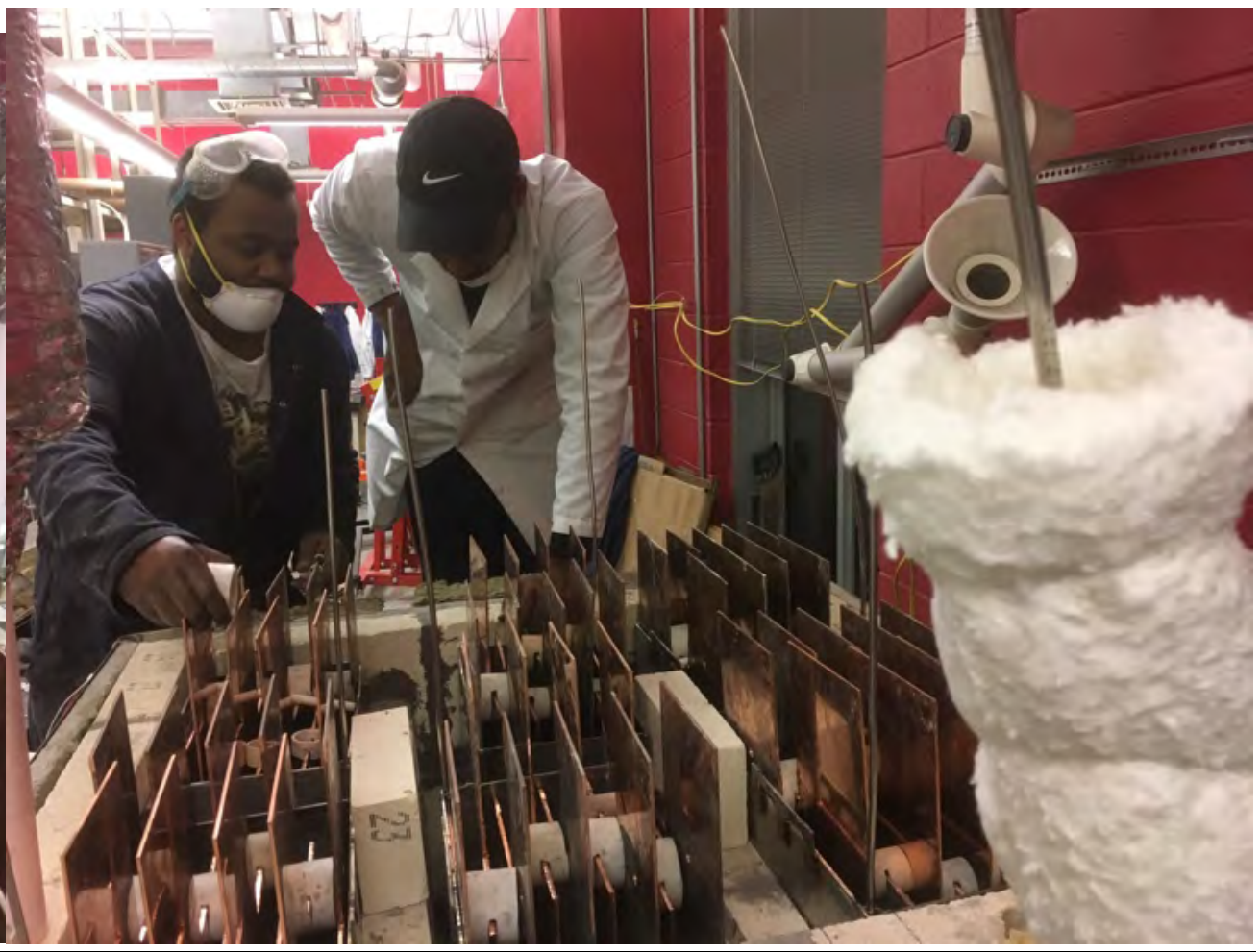

Left: Lithium carbonate electrolyte addition to the Genesis Device™. Right: Genesis device ready for final closure & electrical hookup.

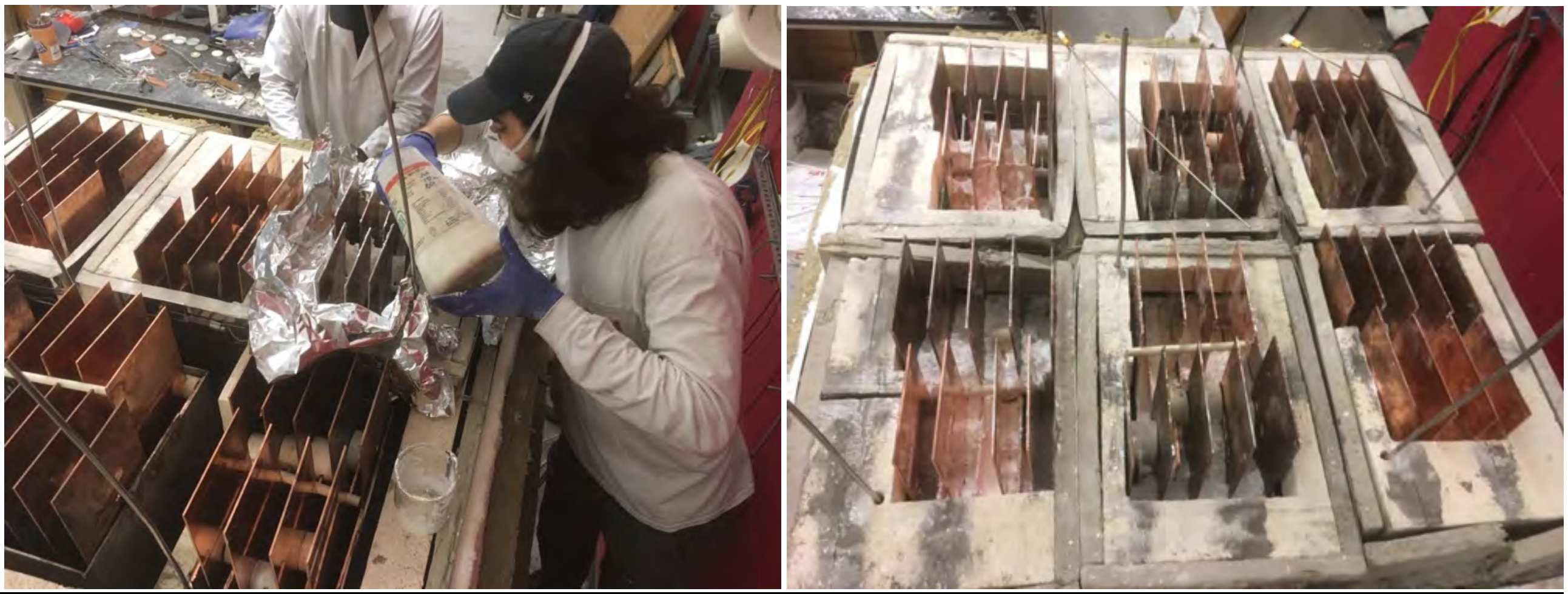

Final completion and sealing the Genesis Device with mineral wool and then R-30, sealing with an additional top dura blanket layer of insulation, and finally the addition of one of several thermocouples.

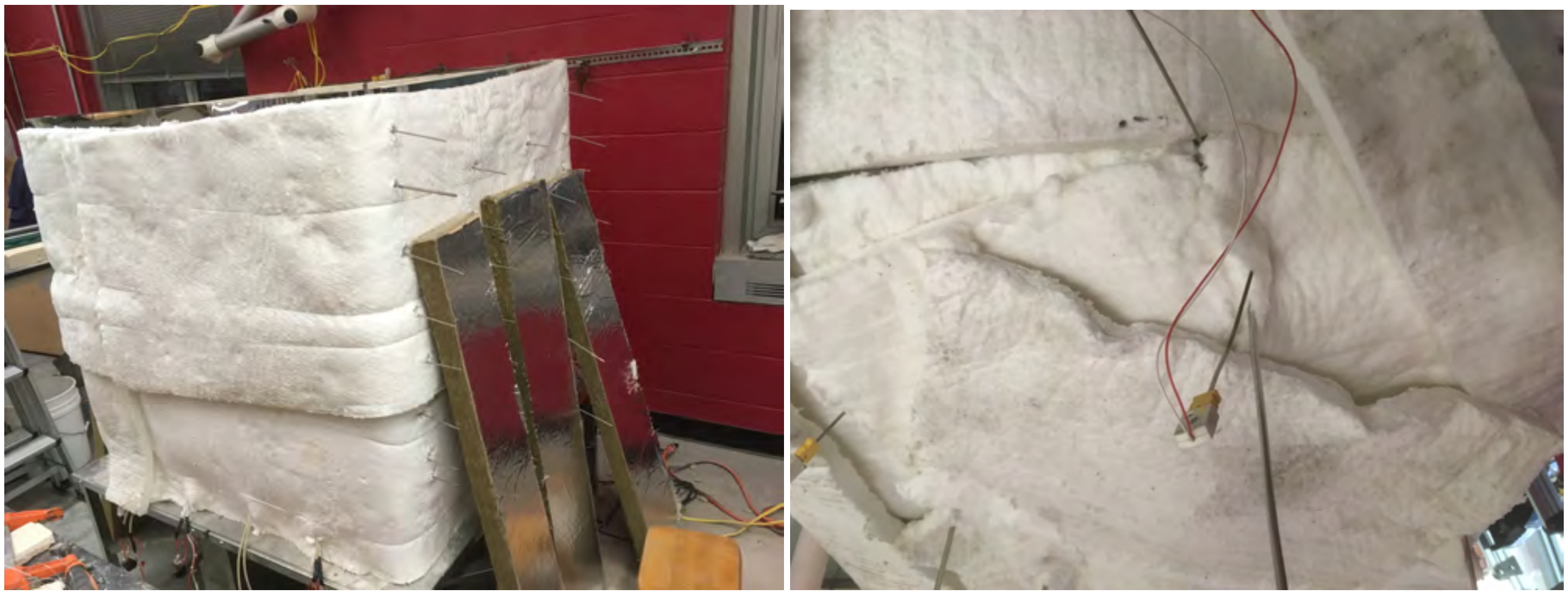